\documentclass[fleqn,usenatbib]{mnras}

\usepackage{newtxtext,newtxmath}

\usepackage[T1]{fontenc}

\usepackage{graphicx}	
\usepackage{amsmath}	
\usepackage{mathrsfs}
\usepackage[flushleft]{threeparttable}
\usepackage{siunitx}
\usepackage{booktabs, caption}
\usepackage{float}
\usepackage{pdflscape}
\usepackage{ulem}
\usepackage{physics}

\title[PAH emission with JWST]{Polycyclic Aromatic Hydrocarbon Emission in Galaxies as seen with JWST}

{\author[D. Rigopoulou et al.] {
D. Rigopoulou$^{1,2}$,\thanks{E-mail: dimitra.rigopoulou@physics.ox.ac.uk}
F. R. Donnan$^{1}$,
I. Garc\'ia-Bernete$^{1}$,
M. Pereira-Santaella$^{3}$,
A. Alonso-Herrero$^{4}$,
\newauthor R. Davies$^{5}$, L. K. Hunt$^{6}$, P. F. Roche$^{1}$, T. Shimizu$^{5}$
\\
$^{1}$ Department of Physics, University of Oxford, Keble Road, Oxford OX1 3RH, UK \\
  $^{2}$ School of Sciences, European University Cyprus, Diogenes street, Engomi, 1516 Nicosia, Cyprus\\
   $^3$ Instituto de F\'isica Fundamental, CSIC, Calle Serrano 123, 28006 Madrid, Spain\\
   $^{4}$ Centro de Astrobiolog\'{\i}a (CAB), CSIC-INTA, Camino Bajo del  
Castillo s/n, E-28692 Villanueva de la Ca\~nada, Madrid, Spain\\
$^5$Max-Planck-Institut fur extraterrestrische Physik, Postfach 1312, D-85741 Garching, Germany\\
$^6$INAF, Osservatorio Astrofisico di Arcetri, Largo E Fermi 5, 50125 Firenze, Italy\\
}
\date{Accepted 2024 June 11. Received 2024 June 7; in original form 2024 May 13}

\pubyear{2024}

\begin{document}
\label{firstpage}
\pagerange{\pageref{firstpage}--\pageref{lastpage}}
\maketitle

\begin{abstract}
We present a systematic study of mid-infrared spectra of galaxies including star-forming galaxies and Active Galactic Nuclei observed with
JWST MIRI-MRS and NIRSpec-IFU. We focus on the relative variations of the 3.3, 6.2, 7.7, 11.3, 12.7 and 17 $\mu$m Polycyclic Aromatic Hydrocarbon (PAH) features within 
spatially resolved regions of galaxies including NGC\,3256, NGC\,7469, VV\,114, II\,Zw96 and NGC\,5728. Using theoretical PAH models and extending our earlier work, we introduce a new PAH diagnostic involving the 17 $\mu$m PAH feature. To determine the drivers of PAH band variations in galaxies, we compare observed PAH spectral bands to predictions from theoretical PAH models.
We consider extinction, dehydrogenation and PAH size and charge as possible drivers of PAH band variations.
We find a surprising uniformity in PAH size distribution among the spatially resolved regions of the galaxies studied here, with no evidence for preferential destruction of the smallest grains, contrary to earlier findings.  Neither extinction nor dehydrogenation play a crucial role in setting the observed PAH bands. Instead, we find that  
PAH charge plays a significant role in PAH inter-band variations. We find a tight relation between PAH charge and the intensity of the radiation field as traced by the [NeIII]$/$[NeII] maps. In agreement with recent JWST results, we find a predominance of neutral PAH molecules in the nuclei of Active Galaxies and their outflows. Ionised PAHs are the dominant population in star-forming galaxies. We discuss the implications of our findings for the use of PAHs as ISM tracers in high redshift galaxies.

\end{abstract}

\begin{keywords}
Infrared: Galaxies -- Infrared: ISM -- Galaxies: active
\end{keywords}



\section{Introduction}

With strong emission features at 3.3, 6.2, 7.7, 8.6, 11.3, 12.7, 16.4 and 17 $\mu$m Polycyclic Aromatic Hydrocarbons (PAH; e.g. \citealp{leger84}) are prominent in the mid-infrared (mid-IR) spectra of a variety of objects, including
H\rm{II} regions, Planetary Nebulae (PNe), post-Asymptotic Giant Branch (AGB) objects, Young Stellar Objects (YSOs), the diffuse
ISM, galaxies and Active Galactic Nuclei (AGN). PAH molecules are thought to consist of 20-1000 carbon atoms 
and, are considered an abundant and energetically important component of interstellar dust, contributing
$\sim$ 5--20\% of the total infrared emission from
a galaxy \citep{smith07}.

The ubiquity of PAHs in astrophysical objects and environments provides a useful diagnostic
of the physical conditions in these sources.  
From an extra-galactic point of view, the
strengths of the 7.7, 6.2 and 11.3 $\mu$m features have been proposed as
tracers of star-formation (e.g., \citealp{rigo99, peet04, ds10}). It is also known that these tracers might be affected
by global galaxy parameters such as the metallicity
\citep{engel08, hunt10}. The diagnostic power of PAH features to trace star formation even in high-z galaxies has been investigated with Spitzer (e.g. \citealp{huang07, pope13, riech14, shipl16}), and more recently with  the James Webb Space Telescope (JWST, e.g., \citealp{spilker23}).
Broadband surveys with the JWST have relied on PAH emission to uncover numerous high-z mid-IR luminous sources (e.g. \citealp{shen23, kirkpatrick23, shivaei24}).

The picture is more complicated for galaxies that host AGN. Such galaxies tend to have low 6.2 and
11.3 $\mu$m
PAH equivalent width (EQW) (e.g., \citealp{rigo99, tran01, sturm00,
aah14}) due to the presence of significant hot
dust continuum and also because the hard AGN photons may destroy the small
PAH molecules (e.g., \citealp{roche91, voit92, siebenmorgen04}).  Ground-based observations have detected
the 11.3 {$\mu$m} PAH feature at distances of hundreds to tens of
parsecs from the AGN (e.g. \citealt{honig10, ram14, igb19,jens17,Esparza-Arredondo18}). Recent JWST observations have provided new insights into the impact of the central black hole on the PAHs suggesting that the AGN has a significant impact on the properties of PAH in the innermost regions  ($\sim$100\,pc) (e.g., \citealp{igb22b,armus23, lai23}) but also on larger scales (e.g., \citealp{igb24a}).

The central wavelength, shape and intensity of each PAH band is known to vary \citep{peet04}. Such variations have been attributed to changes in the structure and properties of PAH molecules under the influence of extreme astrophysical environments. Since each PAH band corresponds to a different vibrational mode, the ratio between different bands can be used as a diagnostic of PAH properties. The 3.3 $\mu$m PAH feature is due to C-H stretching, the 6-9 $\mu$m features originate from C-C stretching whereas the 11.3$\mu$m feature is due to C-H out-of plane bending. \cite{allam99} and \cite{kimsay02} found that C-C modes are intrinsically weaker in neutral PAHs compared to ionised PAHs. As a result, the 6-9$\mu$m features are more prominent in ionised PAHs compared to the 3.3 and 11.3 $\mu$m bands. Hence, the ratio between the C-C and C-H feature intensities provides a good way of probing the charge of PAHs which ultimately relates to the physical conditions of the environment where the emission originates.

\citet{bregtem05} studied the variation
of the 7.7$/$11.3 ratio in reflection nebulae and found a correlation of 
this band ratio with the ratio G$_{\rm 0}$/n$_{\rm e}$ between the integrated
intensity of the UV field, G$_{\rm 0}$, and the electron density,
n$_{\rm e}$. However, \citet{smith07} concluded that this ratio is
relatively constant among pure starburst galaxies but varies by a
factor of 5 among galaxies having a weak AGN. 
They interpret this effect as a selective destruction of the smallest
PAHs by the hard radiation arising from the accretion disk, ruling out
an explanation in terms of ionization of the molecules, in these
particular environments. More recently, \citet{igb22a} found that PAH ratios in  
AGN-dominated systems are consistent with emission from neutral species.

In \cite{rigo21} we used theoretically computed PAH spectra and investigated how the PAH molecular size (parameterized 
by the number of carbons N$_{\rm c}$), the PAH charge (ionised
vs. neutral) and the average energy of the photons impinging on PAH impact the 
relative intensities of each PAH band.
PAH molecules
of different sizes and ionization states vary widely in
the efficiency with which they emit at different bands
\citep{dl07}. However, the underlying mechanism
that governs such variation is not yet fully
understood.

In the present work we wish to use the theoretical PAH grids that were developed in \citet{rigo21} and, through the analysis of JWST spatially resolved observations of nearby galaxies, assess the use of PAH band ratios 
as diagnostics of the physical conditions of the ISM in galaxies.
The previous considerations stress the diversity of the possible interpretation of the mid-IR feature variations in galaxies.
We therefore, need to identify the main physical processes controlling the
PAH bands and use them as diagnostic tools. This 
paper is arranged as follows.
In Section \ref{sec:models} we present a brief summary of the theoretical PAH spectra which have been computed
using Density Functional Theory (DFT) and the ensuing PAH grids. We add one more diagnostic grid employing the PAH 17$\mu$m band which was not widely detected with Spitzer.
In Section \ref{sec:obs} we describe the JWST 
observations of the nearby galaxies that  are used in the present work. We discuss the fits of the mid-IR spectra and measure the intensities of
the various PAH features. 
Section \ref{sec:PAH-var} investigates the impact on spatial PAH band variations of the extinction, charge, molecular size and structure (the latter quantified through the H$/$C ratio) and, finally the presence of the AGN. Discussion and Conclusions follow in Sections \ref{sec:disc} and \ref{sec:conclude}, respectively.

\section{Theoretical PAH models}
\label{sec:models}
\subsection{PAH molecules}
\label{sec:dft} 
In astrophysical settings, when PAH molecules are exposed to UV radiation they absorb photons whose energy is  swiftly transferred to 
the vibrational levels and the molecules proceed to an
excited state. Following excitation, molecules will cool radiatively by emitting IR photons at specific frequencies corresponding to their vibrational modes \citep{allam99, baktil94}. This so-called IR fluoresence is the cornerstone of theoretical models of interstellar PAH emission and can be well approximated by the product of the
absorption cross section convolved with the Planck function, as proposed by \citet{leger84}.
The intrinsic spectral characteristics of PAHs, such as their vibrational frequencies, can subsequently be computed with a variety of techniques as outlined in \citet{tiel08}.
In \citet{rigo21} we used Density Functional Theory (DFT) to compute theoretical spectra of PAH molecules of different sizes (parameterised by the number of C atoms).
In summary, a detailed account of the PAH molecules used in our study was presented in \citet{Kerkeni22} and in Appendix~A of \cite{rigo21}. In this earlier work we examined  the impact 
 of PAH size, charge and hardness of the radiation field on specific PAH bands. We focused primarily on the 3.3, 6.2, 7.7, 11.3 $\mu$m PAH bands whose ratios were subsequently compared to Spitzer-IRS observations corresponding to {\it integrated} measurements in local galaxies. 

With its unprecedented sensitivity and unrivaled spatial resolution JWST offers the possibility to revisit inter-band PAH variations in a spatially resolved manner for a range of nearby galaxies. 
In the present work we revisit the grids presented in \citet{rigo21} and introduce new grids involving the 17 $\mu$m group of PAH which is now easily accessible with JWST and discussed in below.

\subsection{PAH Grids}
Using spectra of theoretically derived PAH molecules with varying numbers of carbons, charge and, exposed to radiation fields of different intensity we can build diagnostic grids that can be directly compared to observations. In \cite{rigo21} we examined the 11.3$/$7.7 vs 6.2$/$7.7 and 11.3$/$3.3 ratios and their ability to trace the charge, the energy of the photons impinging on and the size of PAHs. Here we add one more diagnostic ratio involving the 17$\mu$m PAH. A number of sub-bands at 16.4, 17.0, 17.4 and 17.9 $\mu$m collectively contribute to the 17$\mu$m PAH band which is throught to originate from C-C bending of large PAH molecules (e.g. \citealp{ricca10}).
A detailed analysis of the diagnostic power of the 17$\mu$m feature, in particular, the impact of the number of carbons and the charge on the intensity of the feature is discussed in  Appendix A. 
\begin{figure}
	\includegraphics[width=\columnwidth]{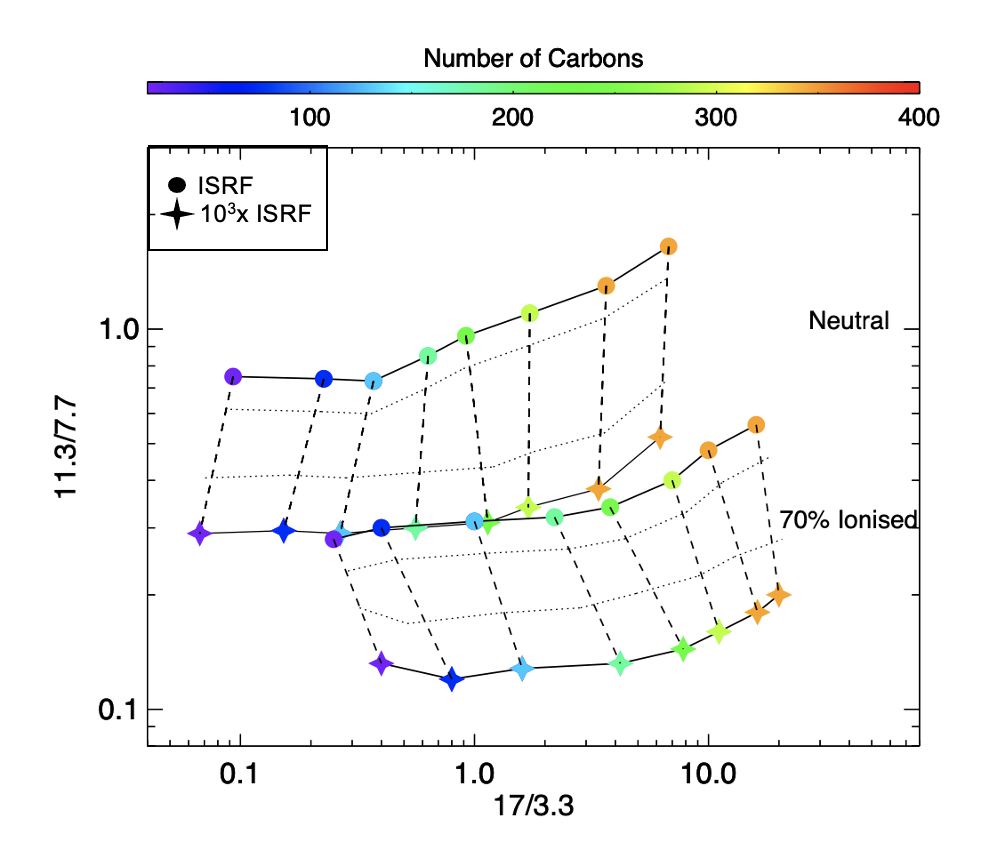}
    \caption{The 17$/$3.3 vs 11.3$/$3.3 PAH ratio plot from DFT spectra exposed to radiation fields with a range of energies from the ISRF (filled circles)
to 10$^{3} \times$ ISRF (stars). The top grid represents neutral PAHs, the bottom grid represents 70\% ionised and 30\% neutral PAHs.}
    \label{fig:example_figure}
\end{figure}

In Figure 1 we show a new diagnostic plot employing the 17 $\mu$m PAH complex (details of the PAH computations that were used in constructing the grids are given in Appendix A) and the 3.3 $\mu$m PAH. This ratio could be a good probe of the PAH size distribution since both bands trace the small and large ends of the PAH size distribution (e.g., \citealp{schutte93}). Figure 1 shows two
grids corresponding to pure neutral PAHs (top) and to a mixture of 30\% neutral and 70\% ionised PAHs (bottom). In each grid the top line (filled circles) corresponds to PAH molecules exposed to the interstellar radiation field (ISRF)\footnote{the
interstellar radiation field (ISRF) in the solar neighbourhood, corresponds to an average photon energy of 12.4 eV.} while the bottom line (filed stars) correspond to a radiation field of 10$^{3}\times$ ISRF.
Together with the grids presented in \citet{rigo21} they will be compared to PAH ratios of different types of galaxies from new JWST observations (see Section \ref{sec:pah-temp}). The goal is to use PAH band ratios to infer the basic properties and, furthermore, to investigate the impact of the energetics of each galaxy on the PAH population.


\section{The Data}
\label{sec:obs}

To investigate the spatially resolved properties of the PAH bands we use JWST observations of five galaxies: four galaxies with combined  NIRSpec-IFU and MIRI-MRS galaxies and one galaxy with only MIRI-MRS observations. The four galaxies with NIRSpec-IFU and MIRI-MRS observations represent different types of nuclear activity from type 1 Seyfert, to Compton-thick AGN, Starburst
and deeply dust-obscured nuclei.
NGC~7469, NGC~3256, VV114 and IIZw96 come from the Director's Discretionary Early Release Science (ERS) observations ID\,1328 (PI: L. Armus \& A. Evans). Our analysis also includes MIRI-MRS spectroscopic observations for NGC 5728, part of the Cycle 1 GO proposal ID\,1670, (PI: T. Shimizu and R. Davies). NGC 5728 is classified as a Compton-thick AGN with a strong outflow that is impacting the disk of the galaxy. The MIRI-MRS spectra of NGC~5728 have been presented in \cite{igb24a},\citet{igb24c}, \citet{davies24}. 

Maps of the 6.2 $\mu$m PAH emission of NGC~7469, NGC~3256, VV114, IIZw96 and NGC~5728 are shown in Figure 2 together with the apertures used to extract nuclear and circumnuclear spectra in each of them. 

\subsection{Spectroscopic observations}

We used mid-IR (4.9--28.1~$\mu$m) MIRI$/$MRS integral-field spectroscopy data (\citealp{Wright23}). MRS comprises 4 wavelength channels: ch1 (4.9–7.65~$\mu$m), ch2 (7.51–11.71~$\mu$m), ch3 (11.55–18.02~$\mu$m) and ch4 (17.71–28.1~$\mu$m). These channels are further subdivided into 3 sub-bands (Short, Medium and Long). The field of view increases with wavelength: ch1 (3\farcs2 $\times$ 3\farcs7), ch2 (4\farcs0 $\times$ 4\farcs7), ch3 (5\farcs2 $\times$ 6\farcs1), and ch4 (6\farcs6 $\times$ 7\farcs6). The spaxel size ranges from 0\farcs13 to 0\farcs35. The spectral resolution of the MRS varies with wavelength with values R$\sim$3700--1300 (\citealp{Argyriou23}). The nuclear spectra from the different sub-channels were extracted assuming they are point sources ($\sim$0.4$\arcsec$ at 11\,$\mu$m). Detailed investigations based on the MIRI-MRS spectra have also appeared in: NGC~7469 \citep{igb22b, lai22, armus23, zhangho23}, 
VV114 \citep{fd23,rich23}, II\,Zw\,096 \citep{igb24b}.
Analysis of the MIRI-MRS data for NGC~5728 have been presented in \cite{igb24c} and \cite{davies24}.

High spectral resolution (R$\sim$2700) 1.0 – 5.27 $\mu$m
NIRSpec integral-field unit (IFU) spectroscopy of NGC~7469, NGC~3256, IIZw96, and VV114 were been obtained
with the grating/filter pairs G140H (0.97–1.89~$\mu$m), G235H (1.66–3.17~$\mu$m) and G395H (2.87–5.27~$\mu$m) with R$\sim$2700. 
The data were
obtained with the NRSIR2RAPID readout using 18 groups and
a 4-point cycling dither pattern. The field of view of NIRSpec is 3\farcs1 $\times$  3\farcs1, with a spaxel size of 0.1". We extracted spectra using a 0.3'' radius aperture. To account for differences in the sub-channels and ensure that the regions are the same between the various sub-channels additional corrections were applied as detailed in \cite{fd24}. For NIRSpec point-like extractions we corrected the nuclear apertures using PSF wavelength dependent correction factor (calibration point source TYC\,4433-1800-1, Program ID 1128, PI: N. Luetzgendorf; see \cite{mps22, igb22b,igb24b, fd23}). Detailed investigations based on NIRSpec spectroscopy have appeared in NGC~7469 \citep{lai23}, NGC~3256 \citep{mps24}, VV114 \citep{GonzalezAlfonso23, Buiten24}, IIZw96 \citep{igb24b}.

We refer the reader to \cite{Rigby23} for further details on the JWST telescope. For the reduction, we followed the standard JWST pipeline procedure (e.g. \citealt{Labiano16} and references therein) and the same configuration of the pipeline stages described in \citet{igb22b} and \citet{mps22} to reduce the data. We used the pipeline release 1.9.4 and the calibration context 1063. Some hot and cold pixels are not identified by the current pipeline, so we added some extra steps as described in \citet{igb24a} and \citet{mps24} for MRS and  NIRSpec, respectively. 

\begin{figure*}
	\includegraphics[width=\textwidth]{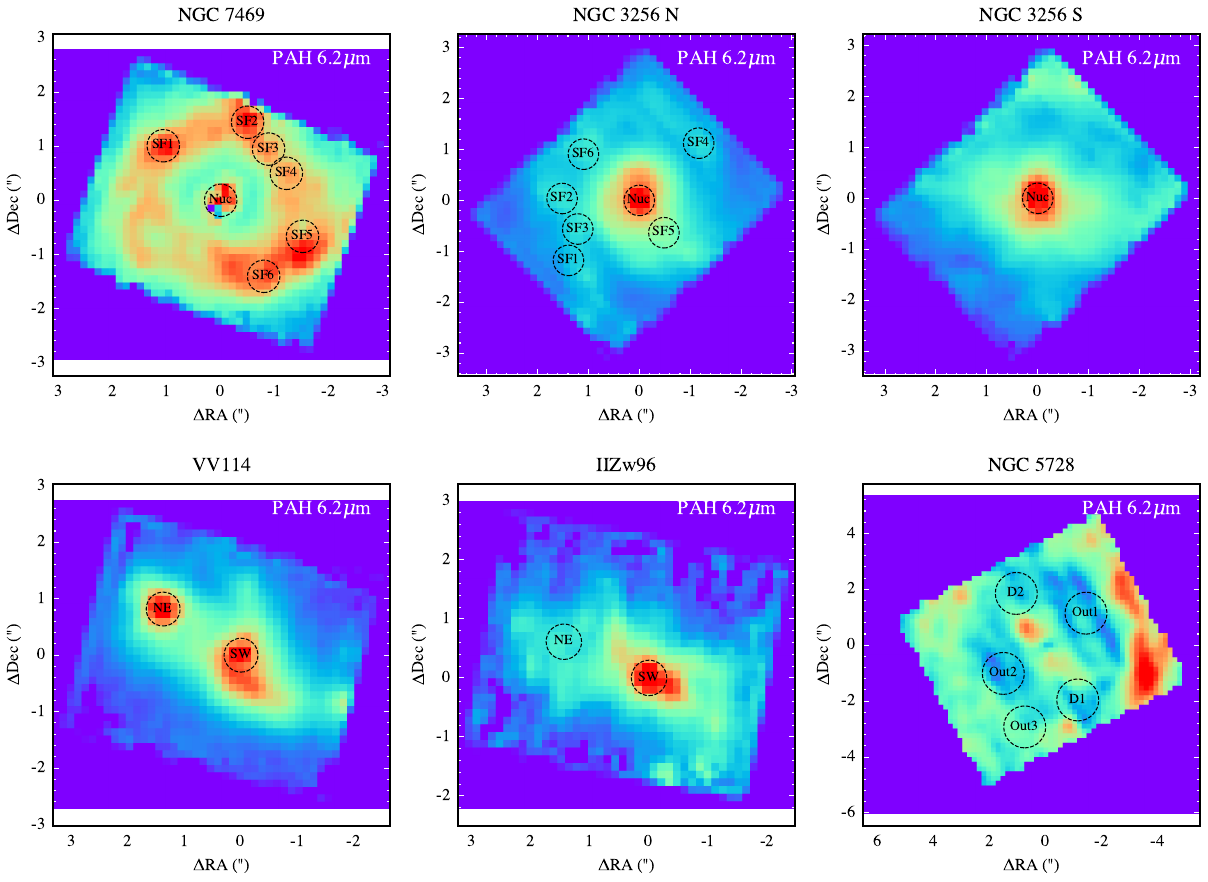}
    \caption{Maps of the 6.2 $\mu$m PAH feature from MIRI MRS data of 6 of the targets in this work. Overlayed are the apertures from \citet{fd24} for NGC 7469, NGC 3256 N, NGC 3256 S, VV114 and IIZw96, all of which have a radius of 0.3". The last panel shows the apertures from \citealp{igb24c} of NGC 5728 which have a radius of 0.75". For NGC5728 the five apertures target regions that are impacted by the AGN outflow.}
     \label{fig:pah-map}
\end{figure*}

In Fig.\ref{fig:pah-map}  we show the positions of the apertures used in this work for each object. We extract six star-forming regions from each of NGC\,7469 and NGC\,3256-N, chosen based on near-IR bright regions. For
NGC 7469, SF2, SF3 and SF4 lie on the redshifted region of the
nuclear outflow (e.g. \cite{igb22b, uviv22, bianchin23}) while the nuclear bar extends (see Fig. 10 of \cite{tds07}) from SF1 to the southwest. For NGC\,3256-S, the aperture is centered on the nucleus. For VV114 there are two apertures centered on the NE and SW nuclei (see \citealp{fd23}). For IIZw96, we define two apertures centered on NE and SW nuclei ( see \citealp{igb24b}). Finally, for NGC\,5728 we define 3 apertures located in the outflow region and two regions located in the disk of the galaxy but in areas impacted by the outflow.
The spectra extracted from each of these apertures have been presented in \citep{fd24} together with a detailed model which employs a wavelength-dependent differential extinction to fit the 1-28 $\mu$m spectra. 

\subsection{Measuring the Intensity of the PAH features}

Measurements of the intensity of the PAH features are subject to uncertainties due to the complexity both of the band profiles but also of the underlying continuum. Unlike emission lines, a large fraction of the energy emitted through the PAH bands is located in the wings of the feature. The wings of the PAH bands can extend beyond the central wavelength of the feature and often blend with the continuum. With contributions from  photospheric emission of old, cool stars on the one side of the spectrum and from stochastically heated small grains on the other (and sometimes 
reprocessed emission from larger grains) building a model for the underlying mid-IR continuum is rather challenging. Such a model must also account for the presence of broad absorption bands of silicate (at 9.7 and 18 $\mu$m) and water and ice features (e.g. \citealp{ dartois07, spoon01, igb24a}). 

In the ISO era (\citealp{kess96}), PAH bands were determined by the spline method (e.g. \citealp{hony01, vermeij02}) which remained oblivious to the complex underlying physical components of the continuum. PAHFIT \citep{smith07} was extensively used to decompose Spitzer-IRS spectra and worked remarkably well for relatively unobscured star-forming galaxies but struggled to model spectra of AGN and$/$or deeply obscured galaxies (such as Luminous and Ultraluminous IR Galaxies, U$/$LIRGs). QUESTFIT \citep{veilleux09} relied on semi-empirical templates of U$/$LIRGs however,  degeneracies were often present in those models (see \citealp{spoon22} for a discussion).

With JWST's superb spatial and spectral resolution we now have an unprecedented view of the PAH bands and the underlying continuum. The enhanced spatial resolution enables detailed studies of nuclear vs. circumnuclear PAH emission of galaxies free-from the effects of dilution which plagued the majority of extragalactic spectra from previous missions. \cite{fd24} presented a new model for fitting the combined NIRSpec and MIRI-MRS spectra of galaxies employing several modified blackbodies to represent dust emission which are then subjected to a wavelength-dependent extinction. The model includes PAH emission features as well as molecular absorption features (in particular water ices at 3 and 6 $\mu$m, CO$_{2}$ at 4.6 $\mu$m and CH at 7 $\mu$m). 

The \cite{fd24} model was used to fit the spectra extracted from each of the apertures shown in Figure \ref{fig:pah-map}. An example of such a spectral fit is shown in Figure \ref{fig:example_spectra}.
The spectral fits for the remaining apertures have been shown in Figures 8 and Appendix D1 of \cite{fd24} while the spectra from the apertures extracted from NGC\,5728 have been shown in \cite{igb24c}.
In Table \ref{tab:pah-tab},  we report fluxes
for the PAH features for the nuclei and the  circumnuclear regions as defined in Fig. \ref{fig:pah-map}. For NGC~5728 the values have been reported in \citealp{igb24c} and are reproduced here for completeness.

\begin{table*}
	\centering
	\caption{Extinction-corrected PAH fluxes} 
	\label{tab:pah-tab}
	\begin{tabular}{lcccccr} 
		\hline
Name & PAH$\lambda$3.3 & PAH$\lambda$6.2 & PAH$\lambda$7.7&PAH$\lambda$11.3&PAH$\lambda$12.7&PAH$\lambda$17$^{1}$\\
& & &(10$^{-14}$ erg cm$^{-2}$ s$^{-1}$)& & & \\
\hline
{\bf NGC3256-N Nuc}&72.52&		301.62&		847.68&	311.44&232.98&	228.90\\
NGC3256 SF1&	2.39&		13.59&	48.89&		14.21&	9.47&		5.45\\	
NGC3256 SF2&	5.01&		19.68&		64.95&		18.32&		12.95&	7.40\\
NGC3256 SF3&	3.86&	19.18&	71.95&	18.94&	14.68&	10.19\\
NGC3256 SF4&	3.07&	15.40&	52.17&	16.57&	11.59&	4.43\\
NGC3256 SF5&	15.64&	57.11&		171.95&	52.28&		27.03&	12.10\\
NGC3256 SF6&3.87&21.02&69.99&19.06&14.93&7.59\\
\hline
{\bf NGC7469 Nuc}&	9.01&	56.81&	225.48&	96.21&		46.58&	121.38\\
NGC7469 SF1&	7.24&		28.94&	83.23&	22.73&		19.51&		10.15\\	
NGC7469 SF2&	6.17&		23.99&		78.73&		17.72&	12.55&		7.73\\	
NGC7469 SF3&	4.35&	24.16&		73.10&		16.95&		11.67&		8.79\\	
NGC7469 SF4&	4.04&		20.72&	61.76&		14.57&		10.41&	8.45\\	
NGC7469 SF5&	4.93&	25.97&	80.50&	20.13&		15.74&		8.33\\
NGC7469 SF6&	5.42&		28.82&79.15&	22.05&		16.49&	8.18\\  
\hline
{\bf Obscured Nuclei} & & & & & \\
VV114NE&35.12&	159.89&520.66&95.62&76.81&	32.97\\
VV114SW&	48.94&	108.31&		268.14&		60.62&	43.55&		67.28\\	
NGC3256-S&	53.18&	189.07&	372.60&		85.13&	50.67&	77.75\\	
IIZw96SW&	8.02&	34.41&	134.84&	19.84&		45.08&		$<$0.02\\
IIZw96NE&	1.04&	3.27&8.29&	3.04&		2.61&		$<$0.06\\	
		\hline
{\bf NGC5728 Nuc$^{2}$}&--&$<$3.47 &26.82 &12.32&--&$<$19.63 \\
NGC5728 Out1&--&1.82&5.16 &3.57 &2.19&1.76 \\
NGC5728 Out2&-- &2.08 &7.56 & 4.54& 2.78& 3.38\\
NGC5728 Out3&-- &3.80 &9.83 &5.41 &3.14 &2.25\\ 
NGC5728 D1&--&2.68&6.82&3.38&2.31&3.62\\
NGC5728 D2&--&2.24&7.26&4.34&2.62&2.42\\
\\
\hline
  
	\end{tabular}
 
$^{1}$: the reported 17 $\mu$m flux is the integral of the 16.45, 17.04 and 17.38 sub-features; $^{2}$: from \cite{igb24c}, no NIRSpec-IFU data are currently available for the this object.
\end{table*}

\section{Spatial Variations of the PAH Band Properties}
\label{sec:PAH-var}

 In the Spitzer era spatially resolved studies of the PAH emission were limited to regions of our Galaxy \citep{galliano08} and a handful of very nearby galaxies 
 (e.g. \citealp{beirao15, mps10, donnelly24}). With low spatial resolution IRS observations, small scale variations vanish when averaged over entire disks and nuclei of galaxies. Therefore, it has been difficult to ascertain what determines PAH band variations and what the real impact of extra-galactic environments -which can be quite different to those in our Galaxy- on PAH bands is.

Using JWST broad band imaging, new insights on spatially resolved PAH band variations have recently been provided (e.g., \citealp{lee23, schin23,ujjwal24}). While such studies can probe the ISM conditions over large scales in nearby galaxies, detailed analysis of the inter-band ratios is difficult especially for the 7.7 \& 8.6 $\mu$m complex where the underlying continuum is also sampled by the broad band filters.

\begin{figure*}
	\includegraphics[width=\textwidth]{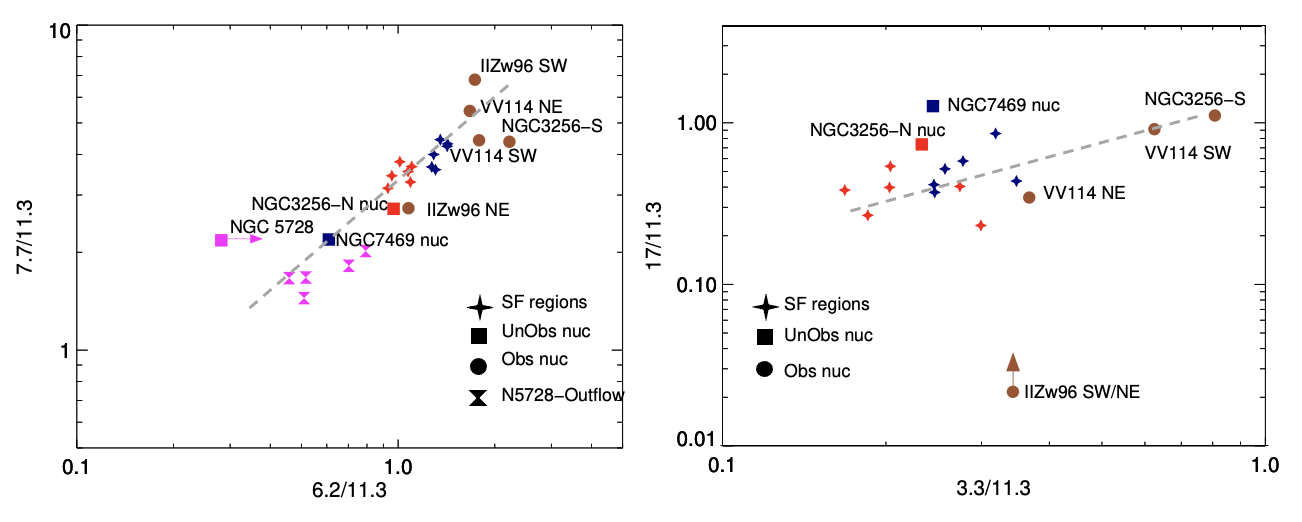}
    \caption{PAH properties for the galaxies studied here,  7.7$/$11.3 vs. 6.2$/$11.3 (left) and 17$/$11.3 vs 3.3$/$11.3 (right). Red and blue stars (squares) correspond to the SF (nuclear) regions in NGC~3256-N and NGC~7469, brown dots represent the obscured nuclei: NGC~3256-S, VV114 NE$/$SW and IIZw96NE$/$SW. The magenta square represents the nuclear aperture for NGC~5728 while the magenta hourglass symbols represent the outflow regions of NGC~5728. Linear fits through both sets of data are shown with the grey dashed lines.}
     \label{fig:pah_values}
\end{figure*}

Figure \ref{fig:pah_values} shows two plots between band ratios of the spatially resolved regions reported in Table \ref{tab:pah-tab}. We employ the brightest bands 3.3, 6.2, 7.7, 11.3, 12.7 and 17 $\mu$m 
and look at the correlations between them. Figure \ref{fig:pah_values} (left) shows that there is a linear correlation between the ratios 6.2$/$11.3 and 7.7$/$11.3. The variations of these ratios spread roughly over an order of magnitude. Likewise, Figure \ref{fig:pah_values} (right) shows the plot between the ratios 3.3$/$11.3 and 17$/$11.3, their values also show a similar variation. Linear fits through both correlations result in similar R values of R$\sim$0.8. We note that for the 6.2$/$11.3 vs 7.7$/$11.3 correlation we have excluded the value for the nucleus of NGC\,5728 as there is an upper limit for the 6.2$\mu$m flux. Likewise, for the 7.7$/$11.3 vs 17$/$11.3 correlation we have excluded the values for the IIZw 96 NE$/$SW apertures as the 17$\mu$m fluxes are upper limits.
The spatially resolved regions of NGC~3256-N, NGC~7469 follow similar trends in both plots while the obscured nuclei, VV114 NE$/$SW, NGC~3256-S and IIZw96 NE$/$SW are often outliers. In what follows we will examine more closely what drives the inter-band variations and how those relate to the physical properties of the galaxies.

There are a number of factors that impact PAH inter-band ratios: the first one is extinction. The deep silicate band originating in the Si-O stretching  mode of amorphous silicates, centred at 9.7 $\mu$m, can impact the band ratios (e.g. \citealp{brandl06, fd23}). Indeed, depending on the depth of the silicate band, the impact on the 11.3 $\mu$m feature can be more significant 
than on the 7.7 $\mu$m. Extinction is also behind the blending of the 8.6 $\mu$m with the 7.7 $\mu$m plateau that is seen sometimes in the spectra of U$/$LIRGs (e.g. \citealp{rigo99, veilleux09}). 
Finally, PAH features and subsequently their ratios, can be impacted by the presence of the 3.0 $\mu$m water ice, 3.4 $\mu$m amorphous hydrocarbon absorption and the 6.2 water ice feature (e.g. \citealp{Sajina09, Spoon04, imanishi06}). We discuss how dust extinction and the presence of ices may impact PAH ratios in \ref{sec:pah-ext}.

The second one is a change in the physical properties of PAHs. Such a change can happen either through modification in the intensity and$/$or shape of the radiation field or, through changes in the PAH size distribution. In addition, changes in the PAH molecule's charge can also impact their physical properties. In \citet{rigo21} (but see also discussion in \citealp{dl21}) the 11.3$/$7.7 ratio was found to be an excellent measure of the fraction of ionised PAH molecules. Likewise, the 6.2$/$7.7 and the 11.3$/$3.3 have been used to track the size distribution of PAHs although for the latter, \citet{rigo21} and \citet{sidhu22} note a dependence on the hardness of the radiation field.
To these two we now add a third one, the 17$/$3.3 ratio (see Section 2 and Appendix A) as a possible ratio to trace PAH sizes. A comparison between these three diagnostic ratios is presented in \ref{sec:pah-temp}.

Finally, dehydrogenation can also play a role in setting the inter-band PAH ratios. Several early laboratory experiments and studies \citep{allam89, andrews16} have shown that dehydrogenation has the same effect on PAH inter-band ratios as ionization. Therefore, examining ratios such as the 11.3$/$6.2 and 12.7$/$11.3 (the latter being sensitive to the presence of duo and trio groups vs. the presence of solo groups of PAHs) would allow us to evaluate whether dehydrogenation plays a crucial role in setting PAH inter-band ratios, as discussed in \ref{sec:pah-hc}.
 
In what follows we investigate each of these factors and the impact they may have on PAH ratios. 


\begin{figure}
	\includegraphics[width=8cm]{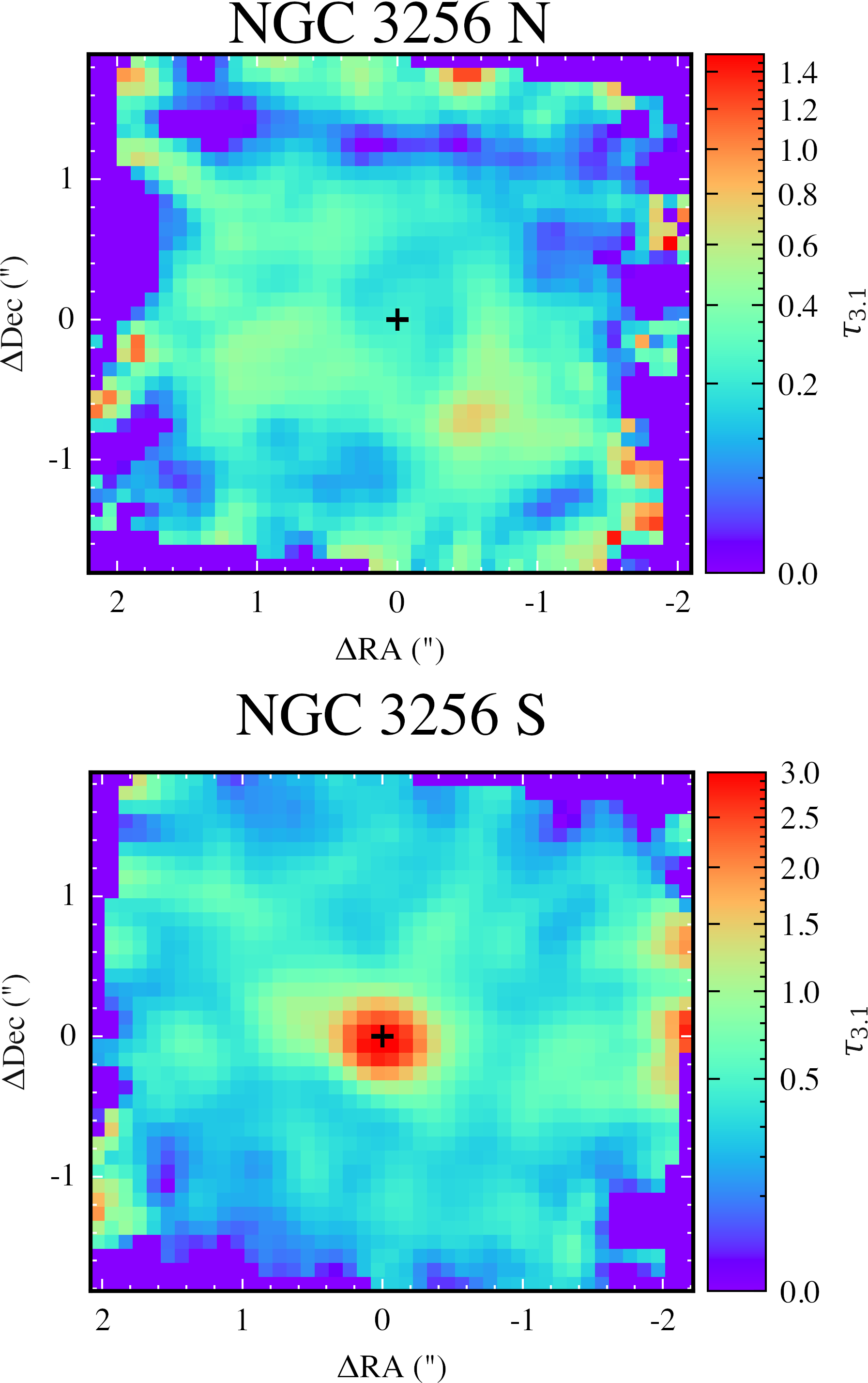}
    \caption{Maps for NGC 3256-N (upper) and NGC 3256-S {\bf{(lower)}} of the optical depth at 3.1$\mu$m due to absorption from water ice. The apertures used for the nuclei and star-forming regions are shown for reference.}
     \label{fig:IceMaps}
\end{figure}

\begin{figure*}
	\includegraphics[width=\textwidth]{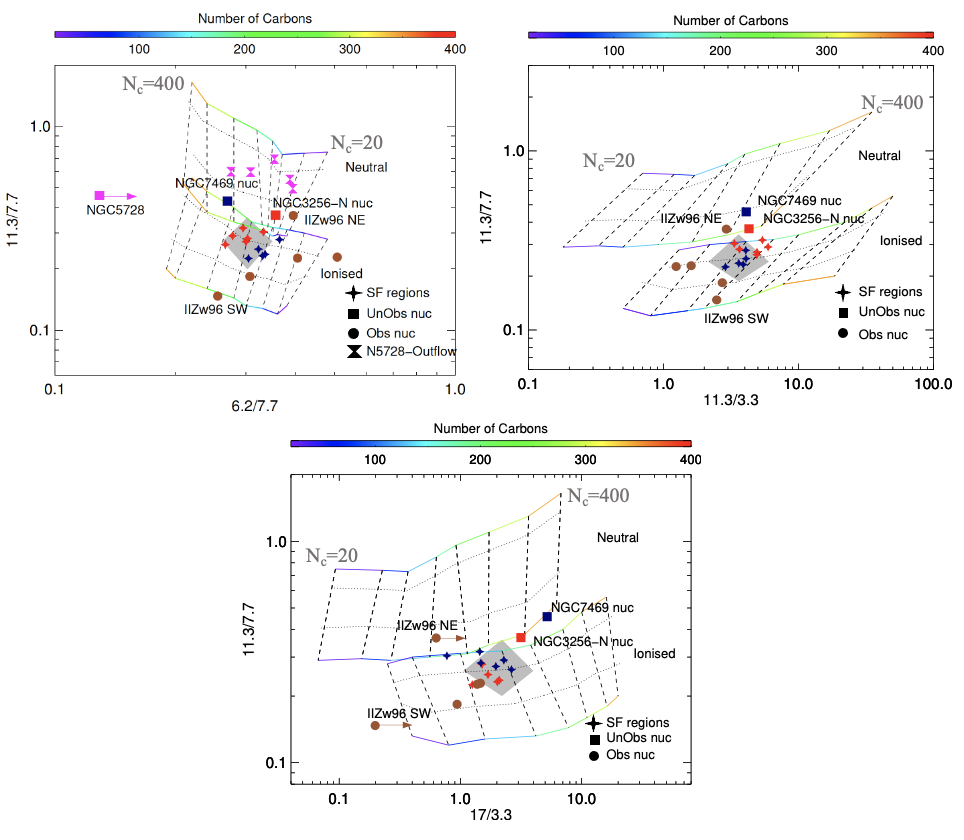}
    \caption{PAH Diagnostic grids involving key PAH bands. From top left clockwise,  11.3$/$7.7 vs. 6.2$/$7.7, 11.3$/$7.7 VS 11.3$/$3.3 and 11.3$/$7.7 vs 17$/$3.3 $\mu$m. The number of carbons (N$_{\rm c}$) going from small to large is indicated in each plot. The grey diamond areas denote the regions occupied by the SF regions defined in Figure \ref{fig:pah-map}. Symbols as in Figure \ref{fig:pah_values}. To avoid overcrowding the Figures with the data points for the obscured nuclei only the values corresponding to IIZw 96 NE$/$SW are explicitly indicated.}
     \label{fig:3rat}
\end{figure*}

\subsection{The effect of extinction on PAH ratios}
\label{sec:pah-ext}
To infer the properties of the PAHs via their ratios, the effect of the extinction must be considered carefully as this can play a crucial role in setting the observed flux ratios \citep[e.g.][]{ahc20}. As discussed in \citet{fd24}, different measures of extinction can be used to correct the measured PAH fluxes. In this work we use the extinction inferred from the HI recombination lines, in particular the ratio of Br$\beta$ (2.62 $\mu$m) to Br$\gamma$ (2.17 $\mu$m), as presented in Table 1 of \citet{fd24}. For the nucleus of NGC~7469, we instead use the extinction inferred from the H$_2$ lines due to contamination of the broad line region towards the HI recombination lines. It is worth noting that PAH are expected to experience lower level of obscuration than H II regions as they are supposed to exist in Photo-dissociation Regions (PDR) on the boundary between ionised and molecular gas (see e.g., Fig. 1 of \citealp{chown23}).

Additionally, the ice absorption at $\sim$ 3 $\mu$m was found to affect the stellar continuum \citep[][]{fd24} appearing spatially extended and not exclusively found in obscured nuclei, unlike the ice absorption at $\sim$ 6 $\mu$m \citep[e.g.][]{igb24a}. In Fig. \ref{fig:IceMaps} we show maps of the optical depth at 3.1 $\mu$m for NGC3256-N and NGC3256-S, the latter hosting an obscured nucleus. To produce these maps, firstly the cube was smoothed with a Gaussian filter with $\sigma =1$ pixel, to improve the signal-to-noise, before a local continuum was set between 2.7 $\mu$m and 3.8 $\mu$m. We then measure the depth of the feature by taking the log of the ratio between the local continuum (f$_{cont}^{\lambda}$) and the measured flux of the feature (f$_{obs}^{\lambda}$), defining 
the optical depth as $\tau_{\lambda}$ = 
-ln( (f$_{obs}^{\lambda}$/f$_{cont}^{\lambda}$).


These maps show the presence of icy grains in the ISM with clear spatial variations. For NGC 3256-N, the nucleus appears to show a relative lack of ices while the star-forming region SF5 shows a high optical depth. In NGC 3256-S, the nucleus has a very deep ice feature \citep[See Fig. 1 of][]{fd24}, with surrounding clumpy/filamentary structures where the ice absorption appears high. Unlike NGC 3256, we do not detect the ice feature within the data of NGC 7469, consistent with the lower extinction towards this target.

With the prevalence of ice absorption at $\sim$ 3 $\mu$m, if one wishes to obtain accurate 3.3 $\mu$m PAH fluxes, this additional extinction must be corrected for. We have therefore corrected the 3.3 $\mu$m PAH fluxes by the measured ice absorption applied to the stellar continuum.
\label{sec:pah-temp}

\subsection{The effect of changes in the physical properties of PAH: charge, size, hardness}
\label{sec:pah-temp}

The peak temperature that a PAH will reach upon absorption of a single UV photon is strongly dependent on the heat capacity of the molecule and therefore its size. In addition, the strength and intensity of the radiation field as well as the ionization state of the PAH
molecule also play a role. 
Following \cite{rigo21} and using the grids presented in Section 2 we can evaluate the impact of size and charge on the PAH band ratios for a variety of sources and regions within galaxies.

The smallest PAH grains have the smallest heat capacities,
and can therefore be stochastically heated to higher
temperatures when excited by single photons. As a result, small PAHs
tend to emit efficiently at shorter wavelengths (3-6 $\mu$m),
while their larger counterparts tend to dominate at
longer wavelengths (10-20 $\mu$m; \citealp{dl07}). A widely used method for probing PAH sizes has been 
the band ratio 6.2/7.7 \citep{dl01, dl21, rigo21,igb22a}.  
Since these two bands are close in wavelength
to each other, the effects of extinction are minimal. In addition, both bands originate from carbon-carbon stretching modes of cations. However, it has been argued (e.g., \citealp{lai20}) that
the range of PAH sizes they probe might be similar.
The combination of a tracer of small PAH molecules to large ones could potentially offer another probe of the size distribution (e.g. \citealp{schutte93}). Therefore, ratios such as 11.3$/$3.3 and 17$/$3.3 can also be used to trace PAH sizes. 

In Fig. \ref{fig:3rat} we show the three size-vs-charge diagnostic plots based on (left-to-right) the 6.2$/$7.7, 11.3$/$3.3 and 17$/$3.3 ratio. In all three diagnostic plots we use the 11.3$/$7.7 ratio as a probe of the charge state of PAH. 
As \cite{igb22a} discuss, the 11.3$/$6.2 could also be used as an alternative probe of the PAH charge. The tracks shown in each diagnostic plot, are for neutral PAHs (top grid) 
and ionised PAHs (bottom grid), starting from small PAHs (N$_{c}$ = 20) to large PAHs (N$_{c}$ = 400). PAH size increases from right to left for the 6.2$/$7.7 diagnostic plot. For the 11.3$/$3.3 and the 17$/$3.3 plots PAH size increases from left to right with the number of carbons (N$_{\rm c}$) given by the colourbar at the top. For each set, PAHs are exposed to radiation fields ranging from ISRF  to 
10$^{3} \times$ ISRF as discussed in \cite{rigo21}. 
In all three diagnostic plots the ISRF increases from top to bottom. 
Therefore, each plot encompasses the combination of all effects that impact the PAH temperature distribution: the intensity of the field, the charge and the PAH size distribution. 

With the full 1.8-28 $\mu$m spectra the entire ensemble of PAHs is accessible to
our study and we can compare the diagnostic power of the three ratios considered, 6.2$/$7.7, 11.3$/$3.3 and 17$/$3.3. We use the spectra from the apertures shown in Figure 2 with the measured PAH intensities reported in Table 1. For the 6.2$/$7.7 plot we include, in addition, spatially resolved data from MIRI-MRS observations of NGC\,5728 \citep{igb24c}. All the PAH intensities have been corrected for extinction following the prescription of \cite{fd24}. To first order, there appears to be a good agreement in the PAH size distribution among the three diagnostic ratios considered here. The SF regions within NGC\,3256-N and NGC\,7469, the nuclear regions of NGC\,3256, NGC\,7469 and the outflow regions of NGC\,5728 
all appear to lie in the middle of the model grids corresponding to PAHs with sizes (ie., number of carbons) in the 150-250 range.
However, the PAH band ratios for the obscured nuclei (NGC\,3256-S, VV114 NE$/$SW and IIZw96 NE$/$SW) show a much wider spread in their values. These trends are evident in all three plots involving each of the 6.2$/$7.7, 11.3$/$3.3 and 17$/$3.3 ratios. We therefore, conclude that all three band ratios are capable of tracing the size distribution of PAHs.

Turning to the 11.3$/$7.7 ratio which probes the charge state of PAHs we find that 
the nucleus of NGC\,7469 (a type 1 Seyfert nucleus) appears to favour more neutral PAH molecules compared to e.g. the nucleus of NGC\,3256-N. Rather surprisingly, we find that the outflow regions in NGC\,5728 contain neutral PAHs as has already been noted in \citealp{igb24c}.
The SF-regions of NGC\,3256 and NGC\,7469 tend to have more ionized PAHs compared to the nuclear regions of the same galaxies. 
Turning to the obscured nuclei, we find that these objects appear to favour quite ionised PAHs more so than the SF-regions of NGC\,3256 and NGC\,7469. 
Fig. \ref{fig:3rat} shows that the highest variation in the ratios involved in the plots is shown in the 11.3$/$7.7 ratio. Indeed, the 11.3$/$7.7 ratio varies by a factor of $\sim$10 whereas the 6.2$/$7.7, 11.3$/$3.3 and 17$/$3.3 vary by factors 2-5. It thus appears that the inter-band variations of PAHs are primarily due to the variation in the fraction of ionized PAHs present. 

\begin{figure}
	\includegraphics[width=\columnwidth]{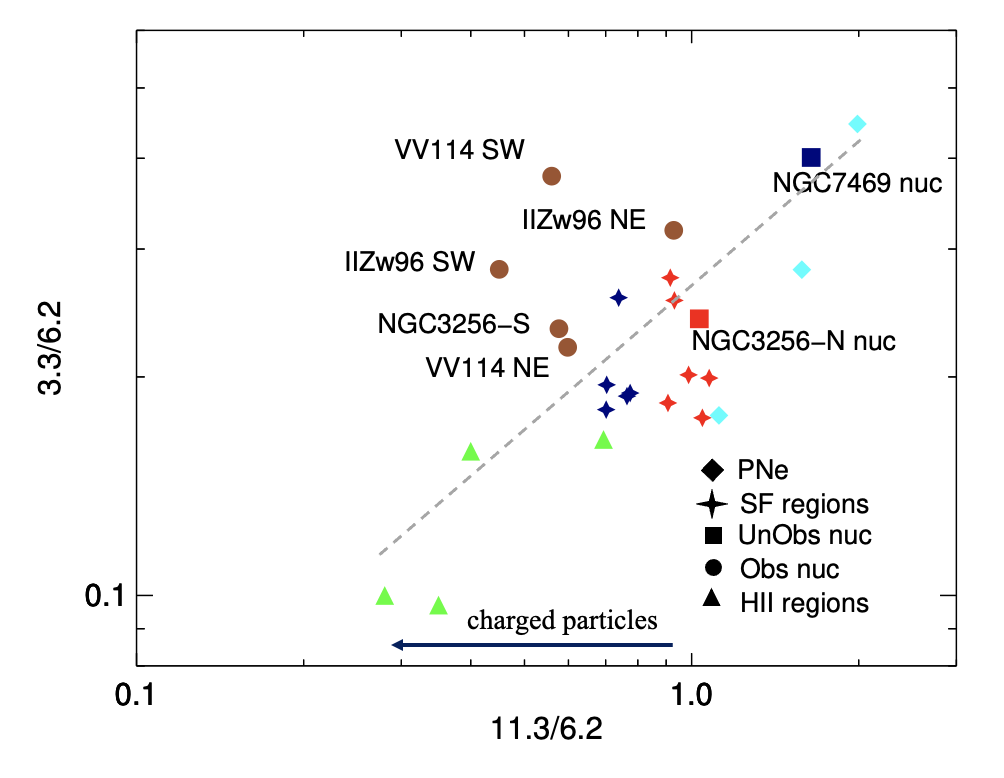}
    \caption{Band strength ratios from spatially resolved regions within galaxies. Symbols as before, in addition data from ISO for HII regions and PNes. Symbols are the same as those used in Figure \ref{fig:pah_values} with the addition of HII regions (green triangles) and PNes (light blue triangles) from \citet{hony01}.}
     \label{fig:charg-rat}
\end{figure}

To further investigate whether PAH molecular size or PAH charge have the biggest impact on PAH ratios, in Fig. \ref{fig:charg-rat} we plot the 3.3$/$6.2 ratio as a function of 11.3$/$6.2 for the spatially resolved SF regions as well as nuclear regions of NGC\,3256, NGC\,7469 and the obscured nuclei shown in Figure 2 and Table 1. We supplement the galaxy data with PAH ratio values for galactic HII regions and Planetary Nebulae (PNe)  taken from \cite{sloan07}. Figure \ref{fig:charg-rat} reveals a very good correlation between the two ratios as evidenced by the linear fit with an R value of R$\sim$0.8 which means that the 11.3$/$3.3 
varies by a relatively small amount ($\sim$3-4)
over the full range of targets and environments.
We interpret this as evidence that the PAH charge state plays a more significant role in setting inter-band ratios than size. Our finding is in good agreement with the results reported by \cite{hony01} and \cite{galliano08} who studied samples of Galactic sources and nearby galaxies with ISO and Spitzer. 
The fact that the inter-band PAH ratios appear to be impacted by the molecule's charge justifies the use of these bands as a tracer of the physical conditions of the emitting regions. We discuss this further in Section \ref{sec:disc}.

\subsection{The impact of the H$/$C ratio}
\label{sec:pah-hc}

Laboratory and theoretical studies of PAHs have investigated the impact the hydrogen-to-carbon ratio (H$/$C) may have on the properties of PAH molecules. However, the influence on the 
C-C$/$C-H ratio (i.e. the ratio of the 6.2 or
7.7 $\mu$m PAH band to the 3.3 or 11.2 $\mu$m PAH band) is hard to discern from the impact that the charge has on these ratios. For instance, small values of the 11.3$/$7.7 ratio are expected for molecules with a small H$/$C ratio (e.g. \citealp{li20}) as well as charged (i.e., ionised) molecules. However, as \citet{allain96a} discuss, only 
 PAHs containing less than N$_{c}\simeq$ 50 carbon atoms can be considerably
dehydrogenated. At the same time, this threshold corresponds
to the minimum size of PAHs that can survive in most
PDRs (e.g., \citealp{allain96b}). Moreover, partially dehydrogenated PAHs are much less stable against photo-destruction compared to other molecules \citep{allain96a}.
 
\begin{figure}
	\includegraphics[width=\columnwidth]{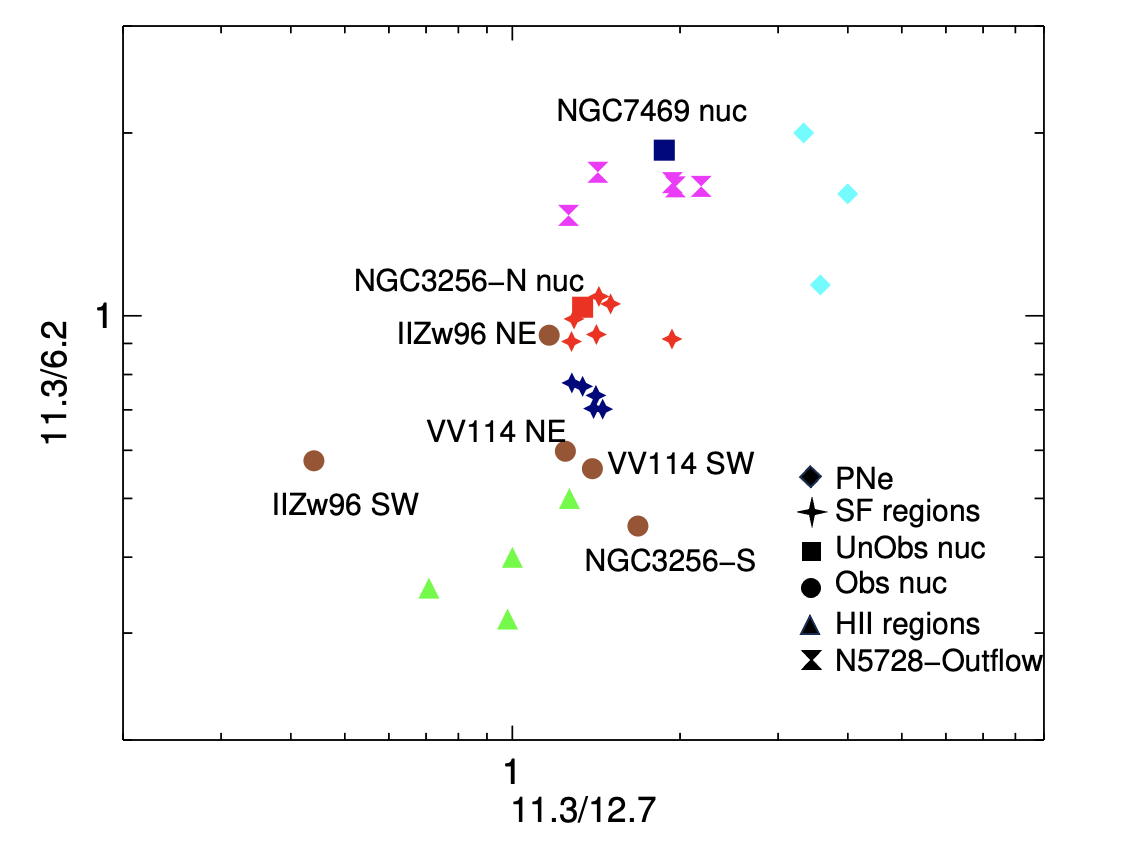}
    \caption{The 11.3$/$6.2 as a function of 11.3$/$12.7 ratio from the spectra corresponding to the regions defined in Figure \ref{fig:pah-map}. Symbols as in Figure \ref{fig:pah_values} with the addition of data for Galactic HII regions (green triangles) and PNes (cyan diamonds) from \citet{hony01}.}
     \label{fig:struc-rat}
\end{figure}

One way to ascertain the role of dehydrogenation  is to examine the relation between solo duo and trio groups (e.g. \citealp{hony01}). Solo$/$duo$/$trio groups refer to the number of CH bonds that are neighbouring a CH unit (see e.g. \citealp{hudgins99}).
There are several inter-band correlations that allow us to assess whether dehydrogenation is important. For instance, when dehydrogenation commences
the number of solo H must increase since  duos and trios
would be converted to solos. As a result we would expect a non-linear relation between 3.3 $\mu$m (originating in all types of CH oscillations) and 11.3 $\mu$m (mostly due to solo oscillations). However, this is not what we observe. Figure \ref{fig:charg-rat} shows a constant (linear) relation between 3.3 and 11.3 over a wide range of sources and environments. In addition, if dehydrogenation were important we would observe a simultaneous decrease in the ratio of 11.3$/$6.2 (decreasing H coverage)
and of the 12.7$/$11.3 ratio (which tracks the conversion of trios and duos into solos (e.g. \citealp{galliano08}).
In Figure \ref{fig:struc-rat} we observe the exact opposite. We conclude therefore that dehydrogenation
has little influence on the observed inter-band relations.

\subsection{The impact of AGN on PAH}
\label{sec:pah-agn}
PAHs have been detected in the vicinity of AGN (e.g. \citealp{honig10, ram14, igb19}) but it
still remains unclear how the Active Nucleus affects band profiles and ratios. While it has
been claimed that PAHs may be destroyed in the vicinity of strong AGN radiation fields (e.g.
\citealp{roche91, voit92, siebenmorgen04, smith07}), PAH emission is detected as close as 10 pc from
the nucleus (e.g., \citealp{honig10}). The advent of JWST faciliates investigations of the PAH properties in the innermost region of AGN
(e.g.,\citealp{igb22b, lai22, fd24, armus23, zhangho23}).
These studies have shown that AGN can have a significant impact on the PAH properties in the innermost $\sim$100 pc. \cite{igb22b} found that central regions of AGN show a larger fraction of neutral PAH molecules compared
to star-forming galaxies as is also evident in Fig. \ref{fig:3rat}. Moreover, in AGN where the outflow is strongly coupled to the host-galaxy (such as in the case of NGC\,5728) PAHs in the outflow are also predominantly neutral. \citealp{igb24c} proposed that the absence of charged PAH in the nuclear and outflow regions of AGN maybe due to their depletion in regions of hard ionizing photons. Upon ionisation,  the structure of PAH molecules changes becoming less stable (e.g. \citealt{voit92}). We can investigate this notion further by looking for clues from the inter-band PAH ratios themselves.

We discussed in Sect. \ref{sec:pah-hc} that the ratio of 11.3$/$12.7 is influenced by the number of large PAHs which are predominantly solos (which is what sets the 11.3 $\mu$m band) whereas the 12.7 $\mu$m band is arising from PAH molecules with edges or open structures. Subsequently, those environments that favour large PAHs (and hence the 11.3 $\mu$m band) also favour neutral PAHs. Likewise, in those regions where open, uneven molecular structures (and hence the 12.7 $\mu$m) dominate, PAHs are predominantly charged. 
Fig. \ref{fig:struc-rat} presents clear evidence
that this is indeed the case: the nucleus of NGC~7469 (a Type 1 Seyfert), the outflow regions of NGC\,5728 and the PNes all have values of the 11.3$/$12.7 ratio which are larger than unity meaning that the 11.3 $\mu$m band is stronger. These are also the regions where the PAH population is predominantly neutral.  
We interpret this as evidence that open PAH molecular structures are less-stable and therefore do not survive in the regions surrounding AGN and$/$or their outflows.
An alternative explanation would involve shielding of the PAH population by H$_{2}$ molecules in the nuclear dusty torus$/$disk as proposed by e.g. \cite{voit92}. However, \citet{igb22b}, do not find evidence that would support H$_{2}$ shielding. The role of H$_{2}$ in shielding PAH molecules will be investigated in detail in a forthcoming paper.

Summarising, in this section we set out to establish what has the biggest impact on the inter-band PAH ratios within spatially resolved regions of galaxies. Our findings indicate that the strength of individual bands are determined by the ionisation stage of PAHs and not by extinction, size or specific molecular structure. Neutral molecules (of all sizes) appear to be more resilient in all environments, from galactic HII regions, PDRs and PNEs, all the way to spatially resolved regions within luminous IR galaxies and AGN.
We discuss this finding and its implications for PAH detections in the next Section.



\section{Discussion}
\label{sec:disc}

\begin{figure*}
	\includegraphics[width=\textwidth]{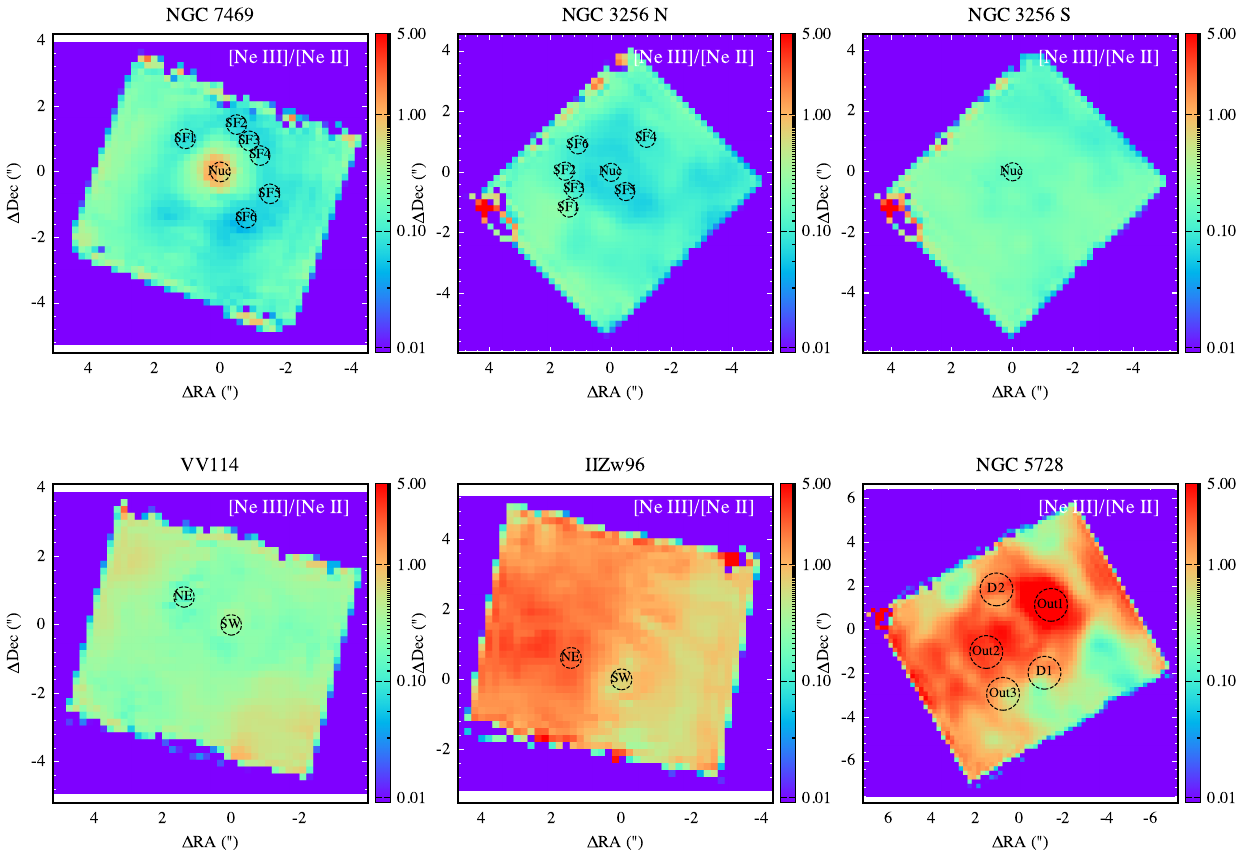}
    \caption{JWST MIRI$/$MRS spatially maps of the [NeIII]/[NeII] emission lines. From top left (clockwise): NGC~7469, NGC~3256-N, NGC~3256-S, VV114, IIZw96 and NGC~5728.The apertures shown in Figure \ref{fig:pah-map} are shown here as well.}
     \label{fig:neii}
\end{figure*}

JWST MIRI/MRS and NIRSpec-IFU observations of nearby galaxies have enabled -- for the first time -- a thorough investigation of the spatial variation and thus properties of PAHs at unprecedented spatial scales. Through the analysis of the PAH ratios from spatially resolved regions in a wide range of environments it appears that the charge of PAHs plays an important role in determining the intensities of the PAH bands. \Citet{dl21} also studied the sensitivity of the PAH bands to the PAH ionization fraction. They concluded that changes in the effective ionization parameter impact the PAH band ratio, in agreement with our findings. 

However, Figure \ref{fig:3rat} reveals a degeneracy between PAH charge and the intensity of the radiation field in the sense that, for some cases, the same 11.3$/$7.7 ratio can be matched with both neutral PAHs in harder radiation fields or charged PAHs in softer radiation fields. To gain more insight into this, in Figure \ref{fig:neii} we plot maps of the [NeIII]/[NeII] ratios. This ratio has been found to provide a good estimate of the intensity of the radiation field in galaxies (e.g. \citealp{thornley00, groves06}). It is evident from this Figure that when the radiation field is very hard (red regions in the maps) the population appears to be dominated by neutral PAHs. The SF regions of NGC~3256 and NGC~7469 are dominated by charged PAHs and the corresponding [NeIII]/[NeII] ratios imply a softer radiation field whereas, the nuclei of NGC~7469 (a type~1 Seyfert), of IIZw96~NE (a deeply obscured AGN, \citealp{igb24b}) and the outflow regions of NGC~5728 are dominated by neutral PAHs. Such observations bring the interesting question of the fate of {\it charged} PAHs in such hostile environments. 

Our analysis in Sections \ref{sec:pah-hc} and \ref{sec:pah-agn} found that the PAH population is self-limiting: PAHs with open structures will not survive the harsh nuclear environments of an AGN or regions of the galaxy disk that are significantly impacted by nuclear outflows. \cite{igb24c}, argued that charged particles are not able to survive in such harsh environments due to Coulomb explosions. This could be a consequence of the increased internal Coulomb forces due the gain or lose of electrons in the molecular system (e.g. \citealt{voit92}). 

Interestingly, our analysis does not find evidence for preferential destruction of small PAH molecules as had been postulated in the Spitzer-era (e.g. \citealp{smith07}). In particular, AGN-dominated environments had been implicated for the destruction  of small PAH grains, perhaps corroborated by the frequent detection of the 11.3 $\mu$m PAH band (e.g., \citealp{ram14}). Instead, we find a predominance of {\it neutral} molecules of {\it all sizes} as those appear to be the most resilient component of the PAH population in the Universe. Figure \ref{fig:charg-rat} indicates that PAH sizes show a relatively small variation among different sources and environments from Galactic HII regions through to the nuclear and circumnuclear regions of Active Galaxies. Interestingly, whereas neutral PAHs are dominant in harsh radiation fields such as those seen in AGN and their outflows, we find a predominance of ionized PAHs in star-forming regions and the obscured nuclei in our study. Galaxies with vigorous star-formation tend to have compact PDRs. In these regions the UV radiation density increases relative to particle (grain) density
impacting the recombination$/$ionisation rate resulting in a higher fraction of charged particles (e.g., \citealp{helou01, mckinney21}).

What do our results imply for the detectability of PAH in galaxies in the high redshift Universe? Clearly, the MIRI$/$MRS detections of the 3.3 $\mu$m PAH in a z$\sim$4.22 lensed galaxy (\citealp{spilker23}) and more recently in a z$\sim$2.4 quasar (\citealp{chen24}) provide clear evidence that PAHs are abundant in high-z galaxies. However, because of the wavelength coverage of MIRI/MRS only the 3.3 $\mu$m PAH is accessible out to redshifts of $\sim$7. The 3.3 $\mu$m PAH band originates in small neutral molecules which, as discussed already, appear to be amongst the more resilient component of the PAH population. 
The detections of the 3.3 $\mu$m PAH in high-z galaxies supports the notion that there is no decline in the 3.3 $\mu$m PAH intensity relative to the total PAH intensity at the high luminosity end in star-forming galaxies (e.g. \citealp{lai20}).
This finding is quite promising for characterising the SF properties of high-z galaxies (e.g. \citealp{shivaei24}).
However, galaxy metallicities have also been found to have an impact on PAH bands (e.g., \citealp{hunt10}). 
Recent JWST observations revealed striking PAH flux ratios in the brightest nucleus of II\,Zw\,096, which might be related to the low metallicity of this source (\citealt{igb24b}). We defer an investigation of the impact of element abundances to the PAH band strengths to a future study.


\section{Conclusions}

\label{sec:conclude}
Based on JWST MIRI/MRS and NIRSpec-IFU spectroscopy of the mid-IR PAH bands of nuclear and circumnuclear regions of local star-forming, AGN$/$Seyfert and obscured galaxies, we investigate what sets the inter-band PAH ratios in these galaxies. 
Our findings are as follows:
\begin{enumerate}
    \item 

We find very good correlation between the main PAH bands 3.2, 6.2, 7.7, 11.3 and 17 $\mu$m of the spatially resolved regions within the galaxies studied here. By employing PAH grids, constructed from theoretically computed PAH models of different size, charge and exposed to radiation fields of different intensities, we confirm that the 6.2$/$7.7, the 11.3$/$3.3 and the 17$/$3.3 are all good tracers of PAH molecular size (albeit the former ratio shows a much larger scatter). The 11.3$/$7.7 (or 11.3$/$6.2) is a good tracer of the PAH charge. 

\item 
We find that extinction plays a role especially towards the shortest wavelength 3.3 $\mu$m PAH band. The presence of ices, especially in the heavily obscured nuclei (such as NGC~3256-S) necessitates the use of extinction corrections in order to obtain accurate 3.3 $\mu$m fluxes.

\item
We confirm a remarkable uniformity in the PAH molecular size (with a variance of at most a factor of 3) over a wide range of sources and environments. We find no evidence for preferential destruction of small-sized PAH molecules, contrary to earlier findings.

\item 
There is no evidence that dehydrogenation plays any role in setting the PAH band ratios. The linear relation between PAH 3.3 $\mu$m (originating in C-H oscillations) and PAH 11.3 $\mu$m (due to solo oscillations) over a wide range of environments is inconsistent with dehydrogenation. 

\item
We find that PAH charge plays a significant role in PAH inter-band variations. There seem to be a close inter-relation between PAH charge and the shape$/$hardness of the underlying radiation field, the latter is traced through spatially resolved [NeIII]$/$[NeII] maps. This result is in agreement with recent findings of a predominance of neutral PAHs in hard radiation fields, e.g. such as those in nuclear regions of AGN$/$Seyferts and their outflows.
On the contrary, SF regions appear to favour more ionised PAHs with the fraction of ionised PAHs reaching 70\%. 

Our work together with recent detections of PAH emission (in particular the PAH 3.3 $\mu$m band) from z$\sim$2-4 galaxies shows that PAH molecules are robust and can be used to trace the SF properties of high-z galaxies out to redshifts of 6.

\noindent
\end{enumerate}

\section*{Acknowledgements}
We thank the anonymous referee for their constructive comments that helped improve the clarity of the manuscript.
DR and IGB acknowledge support from STFC through
grants ST/S000488/1 and ST/W000903/1. FD acknowledges support from STFC through studentship ST/W507726/1. AAH acknowledges support from grant PID2021-124665NB-I00  funded by  
MCIN/AEI/10.13039/501100011033 and by ERDF A way of making Europe. We are grateful to the ERS team for developing their observing program with a zero–exclusive–access period. We thank the Galactic Activity, Torus and Outflow (GATOS) collaboration for many enlightening discussions.
This work is based on observations made with the NASA/ESA/CSA James Webb Space Telescope. The data were obtained from the Mikulski Archive for Space Telescopes at the Space Telescope Science Institute, which is operated  by the Association of Universities for Research in Astronomy, Inc., under  NASA contract NAS 5-03127 for JWST; and from the European JWST archive (eJWST) operated by the ESAC Science Data Centre (ESDC) of the European Space Agency. These observations are associated with programs  1328 and 1670. 


\section*{Data Availability}
 
The JWST data used in this work are publicly available as part of
DD-ERS Program 1328 (PI: L. Armus \& A. Evans), and GO1 - Program 1670
(PI: T. Shimizu \& R. Davies) and downloadable from the MAST archive.
The PAH diagnostic grids are available at {\it https://www.physics.ox.ac.uk/our-people/rigopoulou}



\begin{thebibliography}{99}
\bibitem[\protect\citeauthoryear{Allain, Leach, \& Sedlmayr}{1996a}]{allain96a} Allain T., Leach S., Sedlmayr E., 1996, A\&A, 305, 602

\bibitem[\protect\citeauthoryear{Allain, Leach, \& Sedlmayr}{1996b}]{allain96b} Allain T., Leach S., Sedlmayr E., 1996, A\&A, 305, 616

\bibitem[\protect\citeauthoryear{Allamandola, Hudgins \& Sandford}{1999}]{allam99}
  Allamandola, L.J., Hudgins, D.M., Sandford, S. A., 1999, ApJ
  Lett. 511, 115
\bibitem[\protect\citeauthoryear{Allamandola, Tielens \& Baker}{1989}]{allam89}
 Allamandola, L.J., Tielens, A.G.G.M.,  Barker, J.R., 1989, ApJS 71, 733
\bibitem[\protect\citeauthoryear{Alonso-Herrero et al.}{2020}]{Herrero20} Alonso-Herrero A., Pereira-Santaella M., Rigopoulou D., Garc{\'\i}a-Bernete I., Garc{\'\i}a-Burillo S., Dom{\'\i}nguez-Fern{\'a}ndez A.~J., Combes F., et al., 2020, A\&A, 639, A43. doi:10.1051/0004-6361/202037642

\bibitem[\protect\citeauthoryear{Alonso-Herrero et al.}{2018}]{Herrero18} Alonso-Herrero A., Pereira-Santaella M., Garc{\'\i}a-Burillo S., Davies R.~I., Combes F., Asmus D., Bunker A., et al., 2018, ApJ, 859, 144. doi:10.3847/1538-4357/aabe30

\bibitem[\protect\citeauthoryear{Alonso-Herrero et al.}{2016}]{Herrero16} Alonso-Herrero A., Esquej P., Roche P.~F., Ramos Almeida C., Gonz{\'a}lez-Mart{\'\i}n O., Packham C., Levenson N.~A., et al., 2016, MNRAS, 455, 563. doi:10.1093/mnras/stv2342

\bibitem[\protect\citeauthoryear{Alonso-Herrero et al.}{2014}]{aah14} Alonso-Herrero A., et al., 2014, MNRAS, 443, 2766


\bibitem[\protect\citeauthoryear{Andrews, Candian, \& Tielens}{2016}]{andrews16} Andrews H., Candian A., Tielens A.~G.~G.~M., 2016, A\&A, 595, A23. doi:10.1051/0004-6361/201628819
  
\bibitem[\protect\citeauthoryear{Argyriou et al.}{2023}]{Argyriou23} Argyriou I., Glasse A., Law D.~R., Labiano A., {\'A}lvarez-M{\'a}rquez J., Patapis P., Kavanagh P.~J., et al., 2023, A\&A, 675, A111. doi:10.1051/0004-6361/202346489

\bibitem[\protect\citeauthoryear{Argyriou et al.}{2020}]{Argyriou20} Argyriou I., Wells M., Glasse A., Lee D., Royer P., Vandenbussche B., Malumuth E., et al., 2020, A\&A, 641, A150. doi:10.1051/0004-6361/202037535

\bibitem[\protect\citeauthoryear{Armus et al.}{2023}]{armus23} Armus L., Lai T., U V., Larson K.~L., Diaz-Santos T., Evans A.~S., Malkan M.~A., et al., 2023, ApJL, 942, L37. doi:10.3847/2041-8213/acac66



\bibitem[\protect\citeauthoryear{Bakes \& Tielens}{1994}]{baktil94}
Bakes, E.L.O. \& Tielens, A.G.G.M., 1994, ApJ, 427, 822
\bibitem[\protect\citeauthoryear{Bauschlicher}{2002}]{bausch02}
Bauschlicher, C.W., 2002, ApJ, 564, 782 


\bibitem[\protect\citeauthoryear{Beir{\~a}o et al.}{2015}]{beirao15} Beir{\~a}o P., Armus L., Lehnert M.~D., Guillard P., Heckman T., Draine B., Hollenbach D., et al., 2015, MNRAS, 451, 2640. doi:10.1093/mnras/stv1101

\bibitem[\protect\citeauthoryear{Bianchin et al.}{2023}]{bianchin23} Bianchin M., U V., Song Y., Lai T.~S.-Y., Remigio R.~P., Barcos-Munoz L., Diaz-Santos T., et al., 2023, arXiv, arXiv:2308.00209. doi:10.48550/arXiv.2308.00209
  

\bibitem[\protect\citeauthoryear{Brandl et al.}{2006}]{brandl06} Brandl B.~R., Bernard-Salas J., Spoon H.~W.~W., Devost D., Sloan G.~C., Guilles S., Wu Y., et al., 2006, ApJ, 653, 1129. doi:10.1086/508849

\bibitem[\protect\citeauthoryear{Bregman \& Temi}{2005}]{bregtem05}
Bregman, J., \& Temi, P., 2005, ApJ, 621, 831

\bibitem[\protect\citeauthoryear{Buiten et al.}{2023}]{Buiten24} Buiten V.~A., van der Werf P.~P., Viti S., Armus L., Barr A.~G., Barcos-Mu{\~n}oz L., Evans A.~S., et al., 2023, arXiv, arXiv:2312.01945. doi:10.48550/arXiv.2312.01945

\bibitem[\protect\citeauthoryear{Chen et al.}{2024}]{chen24} Chen Y.-C., Ishikawa Y., Zakamska N.~L., Liu X., Shen Y., Hwang H.-C., Rupke D., et al., 2024, arXiv, arXiv:2403.04002. doi:10.48550/arXiv.2403.04002

\bibitem[\protect\citeauthoryear{Chown et al.}{2023}]{chown23} Chown R., Sidhu A., Peeters E., Tielens A.~G.~G.~M., Cami J., Bern{\'e} O., Habart E., et al., 2023, arXiv, arXiv:2308.16733. doi:10.48550/arXiv.2308.16733

\bibitem[\protect\citeauthoryear{Dartois \& Mu{\~n}oz-Caro}{2007}]{dartois07} Dartois E., Mu{\~n}oz-Caro G.~M., 2007, A\&A, 476, 1235. doi:10.1051/0004-6361:20077798

\bibitem[\protect\citeauthoryear{Davies et al., submitted}{}]{davies24} Davies, R., Shimizu, T., Pereira-Santaella, M., Alonso-Herrero, A., et al., 2024, A\&A, submitted

\bibitem[\protect\citeauthoryear{Diamond-Stanic \& Rieke}{2012}]{Diamond12} Diamond-Stanic A.~M., Rieke G.~H., 2012, ApJ, 746, 168


\bibitem[\protect\citeauthoryear{Diamond-Stanic \& Rieke}{2010}]{ds10}
 Diamond Stanic, A.M., \& Rieke, G.H., 2010, ApJ, 724, 140
 
\bibitem[\protect\citeauthoryear{D{\'\i}az-Santos et al.}{2007}]{tds07} D{\'\i}az-Santos T., Alonso-Herrero A., Colina L., Ryder S.~D., Knapen J.~H., 2007, ApJ, 661, 149. doi:10.1086/513089

\bibitem[\protect\citeauthoryear{D'Hendecourt \& leger}{1987}]{dhende87}
D'Hendecourt, L.B., \& Leger, A., 1987, A\&A 180, 9

\bibitem[\protect\citeauthoryear{Donnan et al.}{2023}]{fd23} Donnan F.~R., Garc{\'\i}a-Bernete I., Rigopoulou D., Pereira-Santaella M., Alonso-Herrero A., Roche P.~F., Hern{\'a}n-Caballero A., et al., 2023, MNRAS, 519, 3691. doi:10.1093/mnras/stac3729

\bibitem[\protect\citeauthoryear{Donnan et al.}{2024}]{fd24} Donnan F.~R., Garc{\'\i}a-Bernete I., Rigopoulou D., Pereira-Santaella M., Roche P.~F., Alonso-Herrero A., 2024, MNRAS.tmp. doi:10.1093/mnras/stae612

\bibitem[\protect\citeauthoryear{Donnelly et al.}{2024}]{donnelly24} Donnelly G.~P., Smith J.~D.~T., Draine B.~T., Togi A., Lai T.~S.-Y., Armus L., Dale D.~A., et al., 2024, arXiv, arXiv:2402.08123. doi:10.48550/arXiv.2402.08123


\bibitem[\protect\citeauthoryear{Draine \& Li}{2001}]{dl01} Draine, B.T., \& Li, A., 2001, ApJ, 551, 807

\bibitem[\protect\citeauthoryear{Draine \& Li}{2007}]{dl07} Draine, B.T., \&Li, A., 2007, ApJ, 657, 810

\bibitem[\protect\citeauthoryear{Draine et al.}{2021}]{dl21} Draine B.~T., Li A., Hensley B.~S., Hunt L.~K., Sandstrom K., Smith J.-D.~T., 2021, ApJ, 917, 3. doi:10.3847/1538-4357/abff51

\bibitem[\protect\citeauthoryear{Engelbracht et al.}{2008}]{engel08}
Engelbracht, C. W., Rieke, G. H., Gordon, K. D., Smith, J.D., Werner,
M. W., et al., 2008, ApJ, 685, 678

\bibitem[\protect\citeauthoryear{Esparza-Arredondo et al.}{2018}]{Esparza-Arredondo18} Esparza-Arredondo D., Gonz{\'a}lez-Mart{\'\i}n O., Dultzin D., Alonso-Herrero A., Ramos Almeida C., D{\'\i}az-Santos T., Garc{\'\i}a-Bernete I., et al., 2018, ApJ, 859, 124. doi:10.3847/1538-4357/aabcbc


\bibitem[\protect\citeauthoryear{Galliano et al.}{2008}]{galliano08}Galliano, F., Madden, S.C., Tielens, A.G.G.M., Peeters, E., Jones,
A. P., 2008, ApJ, 679, 310

\bibitem[\protect\citeauthoryear{Garc{\'\i}a-Bernete et al., submitted}{}]{igb24c} Garc{\'\i}a-Bernete I., Rigopoulou D., Donnan F.~R., Alonso-Herrero, A., Pereira-Santaella, M., et al., 2024, A\&A, submitted 

\bibitem[\protect\citeauthoryear{Garc{\'\i}a-Bernete et al.}{2024b}]{igb24b} Garc{\'\i}a-Bernete I., Pereira-Santaella M., Gonz{\'a}lez-Alfonso E., Rigopoulou D., Efstathiou A., Donnan F.~R., Thatte N., 2024b, A\&A, 682, L5. doi:10.1051/0004-6361/202348744

\bibitem[\protect\citeauthoryear{Garc{\'\i}a-Bernete et al.}{2024a}]{igb24a} Garc{\'\i}a-Bernete I., Alonso-Herrero A., Rigopoulou D., Pereira-Santaella M., Shimizu T., Davies R., Donnan F.~R., et al., 2024a, A\&A, 681, L7. doi:10.1051/0004-6361/202348266

\bibitem[\protect\citeauthoryear{Garc{\'\i}a-Bernete et al.}{2022b}]{igb22b} Garc{\'\i}a-Bernete I., Rigopoulou D., Alonso-Herrero A., Donnan F.~R., Roche P.~F., Pereira-Santaella M., Labiano A., et al., 2022b, A\&A, 666, L5. doi:10.1051/0004-6361/202244806

\bibitem[\protect\citeauthoryear{Garc{\'\i}a-Bernete et al.}{2022a}]{igb22a} Garc{\'\i}a-Bernete I., Rigopoulou D., et al., 2022a, MNRAS, 509, 4256. 

\bibitem[\protect\citeauthoryear{Garc{\'\i}a-Bernete et al.}{2019}]{igb19} Garc{\'\i}a-Bernete I., Ramos Almeida C., Alonso-Herrero A., Ward M.~J., Acosta-Pulido J.~A., Pereira-Santaella M., Hern{\'a}n-Caballero A., et al., 2019, MNRAS, 486, 4917. doi:10.1093/mnras/stz1003

\bibitem[\protect\citeauthoryear{Gonz{\'a}lez-Alfonso et al.}{2023}]{GonzalezAlfonso23} Gonz{\'a}lez-Alfonso E., Garc{\'\i}a-Bernete I., Pereira-Santaella M., Neufeld D.~A., Fischer J., Donnan F.~R., 2023, arXiv, arXiv:2312.04914. doi:10.48550/arXiv.2312.04914

\bibitem[\protect\citeauthoryear{Gonz{\'a}lez-Mart{\'\i}n et al.}{2013}]{Gonzalez-Martin13} Gonz{\'a}lez-Mart{\'\i}n O., Rodr{\'\i}guez-Espinosa J.~M., D{\'\i}az-Santos T., Packham C., Alonso-Herrero A., Esquej P., Ramos Almeida C., et al., 2013, A\&A, 553, A35. Doi:10.1051/0004-6361/201220382

\bibitem[\protect\citeauthoryear{Groves, Dopita \& Sutherland}{2006}]{groves06} Groves B., Dopita M., Sutherland R., 2006, A\&A, 458, 405

\bibitem[\protect\citeauthoryear{Habing}{1968}]{habing68}
Habing, H. J. 1968, Bull. Astron. Inst. Netherlands, 19, 421

\bibitem[\protect\citeauthoryear{Helou et al.}{2001}]{helou01} Helou G., Malhotra S., Hollenbach D.~J., Dale D.~A., Contursi A., 2001, ApJL, 548, L73. doi:10.1086/318916

\bibitem[\protect\citeauthoryear{Hernan-Caballero et al.}{2020}]{ahc20}
Hernan-Caballero, A., Spoon, H.W.W., Alonso-Herrero, A., et al., 2020,
MNRAS, 497, 4614

\bibitem[\protect\citeauthoryear{H{\"o}nig et al.}{2010}]{honig10}
H{\"o}nig, S., Kishimoto, M., Gandhi, P., et al., 2010, A\&A 515, 23 

\bibitem[\protect\citeauthoryear{Huang et al.}{2007}]{huang07}
Huang, J -S., Rigopoulou, D., Papovich, C.,  et al., 2007, ApJ Lett., 660, 69

\bibitem[\protect\citeauthoryear{Hudgins \& Allamandola}{1999}]{hudgins99} Hudgins D.~M., Allamandola L.~J., 1999, ApJL, 516, L41. doi:10.1086/311989

\bibitem[\protect\citeauthoryear{Hony et al.}{2001}]{hony01}
Hony, S., Van Kerckhoven, C., Peeters, et al., 2001, A\&A 378, 41

\bibitem[\protect\citeauthoryear{Hunt et al.}{2010}]{hunt10}
Hunt, L.K., Thuan, T. X., Izotov, Y. I., Sauvage, M., 2010, ApJ 712, 164

\bibitem[\protect\citeauthoryear{Jensen et al.}{2017}]{jens17}
Jensen, J.J., Honig, S.F., Rakshit, S., et al., 2017, MNRAS 470, 3071

\bibitem[\protect\citeauthoryear{Imanishi, Dudley, \& Maloney}{2006}]{imanishi06} Imanishi M., Dudley C.~C., Maloney P.~R., 2006, ApJ, 637, 114. doi:10.1086/498391

\bibitem[\protect\citeauthoryear{Kerkeni et al.}{2022}]{Kerkeni22} Kerkeni B., Garc{\'\i}a-Bernete I., Rigopoulou D., Tew D.~P., Roche P.~F., Clary D.~C., 2022, MNRAS, 513, 3663. doi:10.1093/mnras/stac976

\bibitem[\protect\citeauthoryear{Kessler et al.}{1996}]{kess96}
 Kessler, M.F., Steinz, J. A.; Anderegg, M. E., et al., 1996, A\&A, 315, 27
 
\bibitem[\protect\citeauthoryear{Kim \& Saykally}{2002}]{kimsay02}
Kim, H-S., \& Saykally, R.J.,  2002, ApJS, 143, 455

\bibitem[\protect\citeauthoryear{Kirkpatrick et al.}{2023}]{kirkpatrick23} Kirkpatrick A., Yang G., Le Bail A., Troiani G., Bell E.~F., Cleri N.~J., Elbaz D., et al., 2023, ApJL, 959, L7. doi:10.3847/2041-8213/ad0b14

\bibitem[\protect\citeauthoryear{Lai et al.}{2023}]{lai23} Lai T.~S.-Y., Armus L., Bianchin M., D{\'\i}az-Santos T., Linden S.~T., Privon G.~C., Inami H., et al., 2023, ApJL, 957, L26. doi:10.3847/2041-8213/ad0387

\bibitem[\protect\citeauthoryear{Lai et al.}{2022}]{lai22} Lai T.~S.-Y., Armus L., U V., D{\'\i}az-Santos T., Larson K.~L., Evans A., Malkan M.~A., et al., 2022, ApJL, 941, L36. doi:10.3847/2041-8213/ac9ebf

\bibitem[\protect\citeauthoryear{Lai et al.}{2020}]{lai20}
Lai, T. S.-Y., Smith, J.D.T., Baba, S., Spoon, H.W.W., Imanishi, M.,
2020, ApJ, 905, 55

\bibitem[\protect\citeauthoryear{Labiano et al.}{2016}]{Labiano16} Labiano A., Azzollini R., Bailey J., et al., 2016, SPIE, 9910, 99102W.

\bibitem[\protect\citeauthoryear{Lee et al.}{2023}]{lee23} Lee J.~C., Sandstrom K.~M., Leroy A.~K., Thilker D.~A., Schinnerer E., Rosolowsky E., Larson K.~L., et al., 2023, ApJL, 944, L17. doi:10.3847/2041-8213/acaaae



\bibitem[\protect\citeauthoryear{Leger \& Puget}{1984}]{leger84}
 Leger, \& Puget, 1984, A\&A, 137, 5
 
\bibitem[\protect\citeauthoryear{Leger, d' Hendecourt and Defourneau}{1989}]{legdefour89}
Leger, A., d' Hendecourt, L., Defourneau, D., 1989, A\&A, 216, 148

\bibitem[\protect\citeauthoryear{Li}{2020}]{li20}
Li A., 2020, Nature Astronomy, 4, 339


\bibitem[\protect\citeauthoryear{McKinney et al.}{2021}]{mckinney21} McKinney J., Armus L., Pope A., D{\'\i}az-Santos T., Charmandaris V., Inami H., Song Y., et al., 2021, ApJ, 908, 238. doi:10.3847/1538-4357/abd6f2

\bibitem[\protect\citeauthoryear{Peeters et al.}{2017}]{peet17}
Peeters, E., Bauschlicher, C. W., Allamandola, L. J., et al., 2017,
ApJ, 386, 198

\bibitem[\protect\citeauthoryear{Peeters, Spoon \& Tielens}{2004}]{peet04}
Peeters, E., Spoon, H. W. W., Tielens, A. G. G. M., 2004, ApJ 613, 986

\bibitem[\protect\citeauthoryear{Peeters et al.}{2002}]{Peeters02} Peeters E., Hony S., Van Kerckhoven C., Tielens A.~G.~G.~M., Allamandola L.~J., Hudgins D.~M., Bauschlicher C.~W., 2002, A\&A, 390, 1089. doi:10.1051/0004-6361:20020773



\bibitem[\protect\citeauthoryear{Pereira-Santaella et al.}{2024}]{mps24} Pereira-Santaella M., Gonz{\'a}lez-Alfonso E., Garc{\'\i}a-Bernete I., Garc{\'\i}a-Burillo S., Rigopoulou D., 2024, A\&A, 681, A117. doi:10.1051/0004-6361/202347942

\bibitem[\protect\citeauthoryear{Pereira-Santaella et al.}{2022}]{mps22} Pereira-Santaella M., {\'A}lvarez-M{\'a}rquez J., Garc{\'\i}a-Bernete I., Labiano A., Colina L., Alonso-Herrero A., Bellocchi E., et al., 2022, A\&A, 665, L11. doi:10.1051/0004-6361/202244725


\bibitem[\protect\citeauthoryear{Pereira-Santaella et al.}{2010}]{mps10} Pereira-Santaella M., Alonso-Herrero A., Rieke G.~H., Colina L., D{\'\i}az-Santos T., Smith J.-D.~T., P{\'e}rez-Gonz{\'a}lez P.~G., et al., 2010, ApJS, 188, 447. doi:10.1088/0067-0049/188/2/447

\bibitem[\protect\citeauthoryear{Pope et al.}{2013}]{pope13}
Pope, A., Wagg, J., Frayer, D., et al., 2013, ApJ, 772, 92

\bibitem[\protect\citeauthoryear{Ramos Almeida et al.}{2014}]{ram14} Ramos Almeida C., Alonso-Herrero A., Esquej P., Gonz{\'a}lez-Mart{\'\i}n O., Riffel R.~A., Garc{\'\i}a-Bernete I., Rodr{\'\i}guez Espinosa J.~M., et al., 2014, MNRAS, 445, 1130. doi:10.1093/mnras/stu1756

\bibitem[\protect\citeauthoryear{Ricca et al.}{2012}]{ricca12}
Ricca, A., Bauschlicher, C. W., Boersma, C., Tielens, A.G.G.M.,
Allamandola, L.J., 2012, ApJ, 754, 75

\bibitem[\protect\citeauthoryear{Ricca et al.}{2010}]{ricca10} Ricca A., Bauschlicher C.~W., Mattioda A.~L., Boersma C., Allamandola L.~J., 2010, ApJ, 709, 42. doi:10.1088/0004-637X/709/1/42

\bibitem[\protect\citeauthoryear{Rich et al.}{2023}]{rich23} Rich J., Aalto S., Evans A.~S., Charmandaris V., Privon G.~C., Lai T., Inami H., et al., 2023, ApJL, 944, L50. doi:10.3847/2041-8213/acb2b8

\bibitem[\protect\citeauthoryear{Riechers et al.}{2014}]{riech14}
 Riechers, D., Pope, A., Daddi, E., et al, 2014, ApJ, 796, 84
 
\bibitem[\protect\citeauthoryear{Rieke et al.}{2015}]{Rieke15} Rieke G.~H., Wright G.~S., B{\"o}ker T., Bouwman J., Colina L., Glasse A., Gordon K.~D., et al., 2015, PASP, 127, 584. doi:10.1086/682252
 
 \bibitem[\protect\citeauthoryear{Rigby et al.}{2023}]{Rigby23} Rigby J., Perrin M., McElwain M., Kimble R., Friedman S., Lallo M., Doyon R., et al., 2023, PASP, 135, 048001. doi:10.1088/1538-3873/acb293

 
\bibitem[\protect\citeauthoryear{Rigopoulou et al.}{2021}]{rigo21}
  Rigopoulou D., Barale M., Clary D.C., Shan X., Alonso-Herrero A., Garc{\'\i}a-Bernete I., Hunt L., et al., 2021, MNRAS, 504, 5287

\bibitem[\protect\citeauthoryear{Rigopoulou et al.}{1999}]{rigo99}
 Rigopoulou, D., Spoon, H. W. W., Genzel, R., et al., 1999, AJ, 118, 2625

\bibitem[\protect\citeauthoryear{Roche et al.}{2007}]{roche07} Roche P.~F., Packham C., Aitken D.~K., Mason R.~E., 2007, MNRAS, 375, 99. doi:10.1111/j.1365-2966.2006.11207.x

\bibitem[\protect\citeauthoryear{Roche et al.}{1991}]{roche91} Roche P.~F., Aitken D.~K., Smith C.~H., Ward M.~J., 1991, MNRAS, 248, 606
 
\bibitem[\protect\citeauthoryear{Roche et al.}{1989}]{roche89}
 Roche, P.F, Aitken, D.K., Smith, C.H., 1989, MNRAS, 236, 485

\bibitem[\protect\citeauthoryear{Sajina et al.}{2009}]{Sajina09} Sajina A., Spoon H., Yan L., Imanishi M., Fadda D., Elitzur M., 2009, ApJ, 703, 270. doi:10.1088/0004-637X/703/1/270

\bibitem[\protect\citeauthoryear{Sales, Pastoriza \&Riffel}{2010}]{sales10}
Sales, D.A., Pastoriza, M. G., Riffel, R., 2010, ApJ, 725, 605

\bibitem[\protect\citeauthoryear{S{\'a}nchez-Garc{\'\i}a et al.}{2022}]{Sanchez-Garcia22} S{\'a}nchez-Garc{\'\i}a M., Pereira-Santaella M., Garc{\'\i}a-Burillo S., Colina L., Alonso-Herrero A., Villar-Mart{\'\i}n M., Saito T., et al., 2022, A\&A, 659, A102. doi:10.1051/0004-6361/202141963

\bibitem[\protect\citeauthoryear{Schinnerer et al.}{2023}]{schin23} Schinnerer E., Emsellem E., Henshaw J.~D., Liu D., Meidt S.~E., Querejeta M., Renaud F., et al., 2023, ApJL, 944, L15. doi:10.3847/2041-8213/acac9e


\bibitem[\protect\citeauthoryear{Schutte et al.}{1993}]{schutte93}
Schutte, W. A., Tielens, A. G. G. M., Allamandola, L. J., 1993,
ApJ, 415, 397 

\bibitem[\protect\citeauthoryear{Shen et al.}{2023}]{shen23} Shen L., Papovich C., Yang G., Matharu J., Wang X., Magnelli B., Elbaz D., et al., 2023, ApJ, 950, 7. doi:10.3847/1538-4357/acc944

\bibitem[\protect\citeauthoryear{Shipley et al.}{2016}]{shipl16}
Shipley, H.V., Papovich, C., Rieke, G.H., Brown, M.J.I., Moustakas,
J., 2016, ApJ, 818, 60

\bibitem[\protect\citeauthoryear{Shivaei et al.}{2024}]{shivaei24} Shivaei I., Alberts S., Florian M., Rieke G., Wuyts S., Bodansky S., Bunker A.~J., et al., 2024, arXiv, arXiv:2402.07989. doi:10.48550/arXiv.2402.07989

\bibitem[\protect\citeauthoryear{Sidhu et al.}{2022}]{sidhu22} Sidhu A., Tielens A.~G.~G.~M., Peeters E., Cami J., 2022, MNRAS, 514, 342. doi:10.1093/mnras/stac1255


\bibitem[\protect\citeauthoryear{Siebenmorgen et al.}{2004}]{siebenmorgen04} Siebenmorgen R., et al., 2004, A\&A, 414, 123

\bibitem[\protect\citeauthoryear{Sloan}{2007}]{sloan07}
Sloan, G. C., Jura, M., Duley, W. W., et al., 2007, ApJ, 664, 1144

\bibitem[\protect\citeauthoryear{Smith et al.}{2007}]{smith07}
Smith, J.D.T., Draine, B. T., Dale, D. A., et al., 2007, ApJ, 656, 770 

\bibitem[\protect\citeauthoryear{Smith et al.}{2004}]{Smith04} Smith J.~D.~T., Dale D.~A., Armus L., Draine B.~T., Hollenbach D.~J., Roussel H., Helou G., et al., 2004, ApJS, 154, 199. doi:10.1086/423133


\bibitem[\protect\citeauthoryear{Spilker et al.}{2023}]{spilker23} Spilker J.~S., Phadke K.~A., Aravena M., Archipley M., Bayliss M.~B., Birkin J.~E., B{\'e}thermin M., et al., 2023, Natur, 618, 708. doi:10.1038/s41586-023-05998-6

\bibitem[\protect\citeauthoryear{Spoon et al.}{2022}]{spoon22} Spoon H.~W.~W., Hern{\'a}n-Caballero A., Rupke D., Waters L.~B.~F.~M., Lebouteiller V., Tielens A.~G.~G.~M., Loredo T., et al., 2022, ApJS, 259, 37. doi:10.3847/1538-4365/ac4989

\bibitem[\protect\citeauthoryear{Spoon et al.}{2004}]{Spoon04} Spoon H.~W.~W., Armus L., Cami J., Tielens A.~G.~G.~M., Chiar J.~E., Peeters E., Keane J.~V., et al., 2004, ApJS, 154, 184. doi:10.1086/422813

\bibitem[\protect\citeauthoryear{Spoon et al.}{2001}]{spoon01} Spoon H.~W.~W., Keane J.~V., Tielens A.~G.~G.~M., Lutz D., Moorwood A.~F.~M., 2001, A\&A, 365, L353. doi:10.1051/0004-6361:20000557

\bibitem[\protect\citeauthoryear{Sturm et al.}{2000}]{sturm00}
 Sturm, E., Lutz, D., Tran, D., et al., 2000, A\&A, 358, 481

\bibitem[\protect\citeauthoryear{Thornley et al.}{2000}]{thornley00} Thornley M.~D., Schreiber N.~M.~F., Lutz D., Genzel R., Spoon H.~W.~W., Kunze D., Sternberg A., 2000, ApJ, 539, 641

 
\bibitem[\protect\citeauthoryear{Tielens}{2008}]{tiel08}
Tielens, A.G.G.M., 2008, ARA\&A, 46, 289 

\bibitem[\protect\citeauthoryear{Tran et al.}{2001}]{tran01}
Tran, Q.D., Lutz, D., Genzel, R., et al.,  2001, ApJ 552, 527

\bibitem[\protect\citeauthoryear{U et al.}{2022}]{uviv22} U. V., Lai T., Bianchin M., Remigio R.~P., Armus L., Larson K.~L., D{\'\i}az-Santos T., et al., 2022, ApJL, 940, L5. doi:10.3847/2041-8213/ac961c

\bibitem[\protect\citeauthoryear{Ujjwal et al.}{2024}]{ujjwal24} Ujjwal K., Kartha S.~S., Krishna R A., Mathew B., Subramanian S., P S.T., Thomas R., 2024, arXiv, arXiv:2401.04061. doi:10.48550/arXiv.2401.04061

\bibitem[\protect\citeauthoryear{Veilleux et al.}{2009}]{veilleux09} Veilleux S., Rupke D.~S.~N., Kim D.-C., Genzel R., Sturm E., Lutz D., Contursi A., et al., 2009, ApJS, 182, 628. doi:10.1088/0067-0049/182/2/628

\bibitem[\protect\citeauthoryear{Vermeij et al.}{2002}]{vermeij02} Vermeij R., Peeters E., Tielens A.~G.~G.~M., van der Hulst J.~M., 2002, A\&A, 382, 1042. doi:10.1051/0004-6361:20011628

\bibitem[\protect\citeauthoryear{Voit}{1992}]{voit92} Voit G.~M., 1992, MNRAS, 258, 841

\bibitem[\protect\citeauthoryear{Wells et al.}{2015}]{Wells15} Wells M., Pel J.-W., Glasse A., Wright G.~S., Aitink-Kroes G., Azzollini R., Beard S., et al., 2015, PASP, 127, 646. doi:10.1086/682281

\bibitem[\protect\citeauthoryear{Werner et al.}{2004}]{wern04}
 Werner, M., Roellig, T.L., Low, F.J., et al., 2004, ApJS, 154, 1

\bibitem[\protect\citeauthoryear{Wright et al.}{2023}]{Wright23} Wright G.~S., Rieke G.~H., Glasse A., Ressler M., Garc{\'\i}a Mar{\'\i}n M., Aguilar J., Alberts S., et al., 2023, PASP, 135, 048003. doi:10.1088/1538-3873/acbe66

\bibitem[\protect\citeauthoryear{Zhang \& Ho}{2023}]{zhangho23} Zhang L., Ho L.~C., 2023, ApJL, 953, L9. doi:10.3847/2041-8213/acea73



\end{thebibliography}





\appendix

\section{The 17 micron PAH band}

\label{sec:appA}
\begin{figure*}
	\includegraphics[width=\columnwidth]{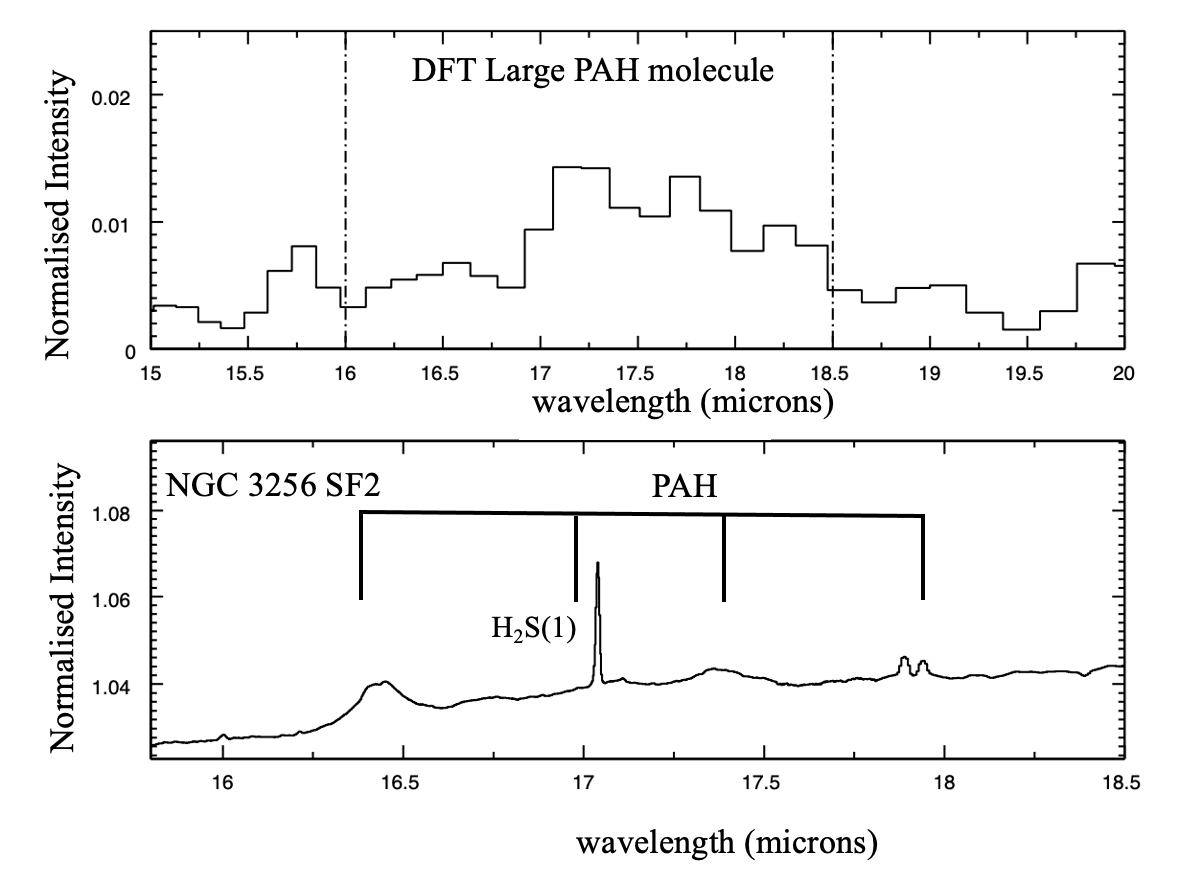}
 \caption{The 17 $\mu$m complex of NGC\,3256-N SF2 star-forming
  region (bottom). Four sub-features at 16.4, 17, 17.4 and 17.9 $\mu$m are indicated,
  they are collectively contributing to the 17 $\mu$m flux. The
  H$_{2}$S(1) line is also indicated. The 17 $\mu$m complex 
  in the spectrum of one of the
 large PAH molecules computed with DFT from \citet{rigo21} (top). The vertical lines
  indicate the region that is used to compute the strength of the
  feature in the DFT PAH molecule.}

     \label{fig:comp17}
\end{figure*}

Longwards of 10 $\mu$m the strongest PAH emission bands include the 11.3, 12.7 and the broad feature at 17 $\mu$m. 
Interestingly, the broad band centered at 17.04 $\mu$m, characteristic of a C-C-C bending mode in neutral PAH, was first identified in the extragalactic source NGC\,7331 \citep{Smith04}.
Observationally, this broad feature is made
up of a group of PAH bands at 16.4, 17.0, 17.4, and 17.9
$\mu$m. Figure \ref{fig:comp17} shows the 17 $\mu$m complex from SF2,
one of the star-forming regions defined in NGC 3256 N (bottom panel), and the PAH-DFT spectrum of the large molecules presented in \citealp{rigo21}. 

\begin{figure}
	\includegraphics[width=\columnwidth]{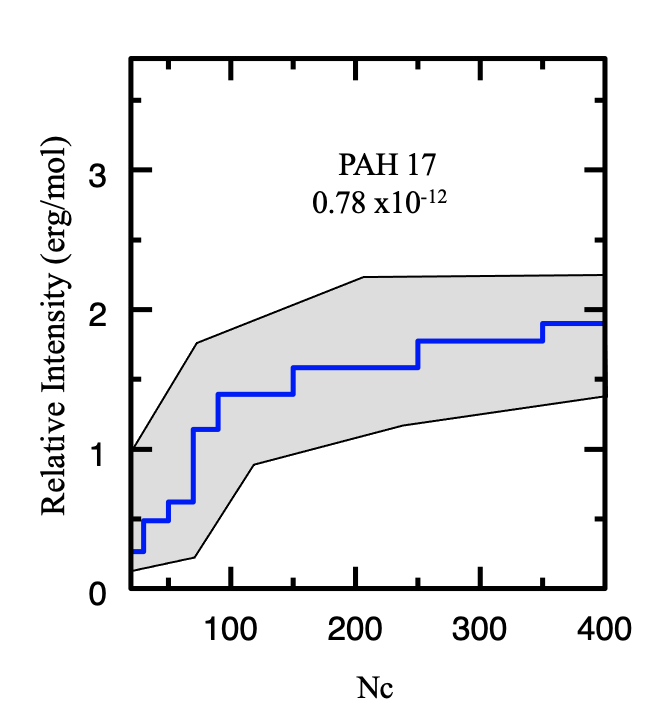}
    \caption{The normalised intensity for the 17 $\mu$m complex for neutral PAH as a function of N$_{c}$, number of carbons. The average value in each bin is represented with the line while the shaded region denotes the spread of values in each bin. Those PAHs with intensity values greater than 3$\sigma$ have been excluded. The intensities have been calculated for the ISRF radiation field.}
    \label{fig:PAH17_size}
\end{figure}

The strength of the 17 $\mu$m complex as a function of the number of carbons, N$_{c}$ is determined following the procedure outlined in \citealp{rigo21}. Figure \ref{fig:PAH17_size} shows the average intensity of the feature together with the 3$\sigma$ spread around this average value for every PAH bin from the smallest to the largest bins (the size distribution of all PAHs used to determine the intensities is shown in Fig. 2 of \citealp{rigo21}). 

\begin{figure}
	\includegraphics[width=\columnwidth]{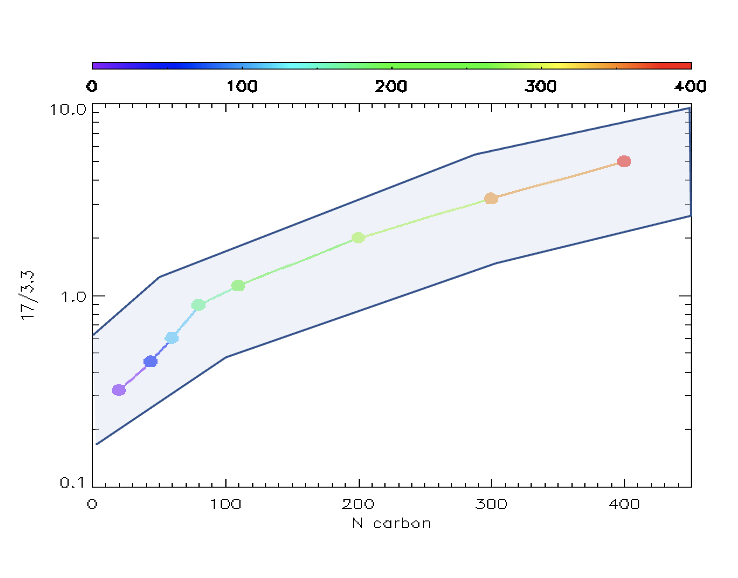}
    \caption{The 17$/$3.3 intensity ratio as a function of N$_{C}$ for neutral PAHs exposed to the Galaxy ISRF radiation. The shaded region corresponds to the spread of values of each individual PAH considered in this study.}
     \label{fig:rat}
\end{figure}

\begin{figure}
	\includegraphics[width=\columnwidth]{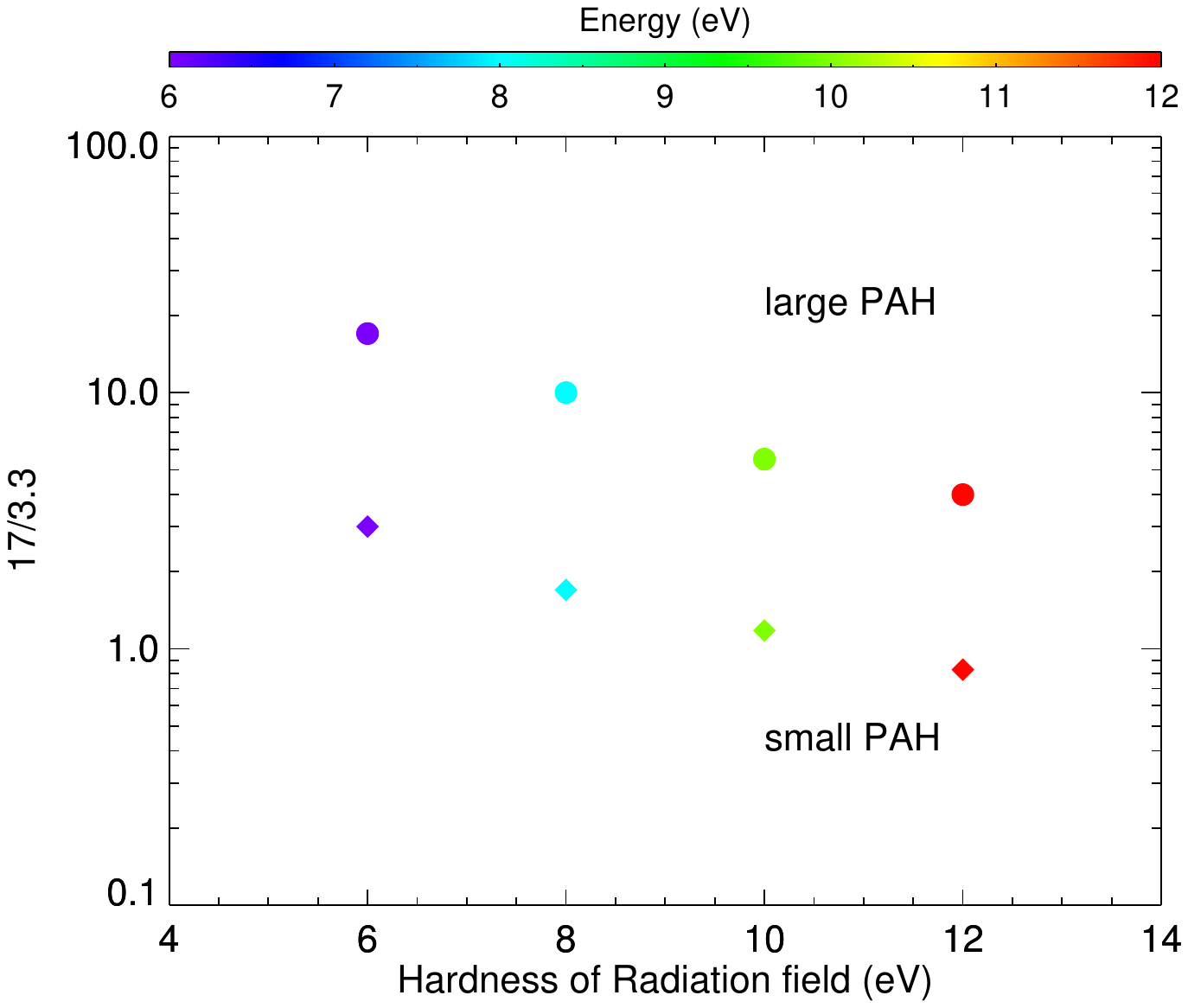}
    \caption{ The 17$/$3.3 intensity ratio calculated for the small bin-averaged molecule (bottom) and the largest bin-averaged (top) neutral PAH molecules as a function of radiation fields with varying energy from 6 to 12 eV. The colour coding corresponds the hardness of the radiation field.}
    \label{fig:pah_hard}
\end{figure}

Since the 17 $\mu$m complex is associated with large neutral PAH molecules e.g. \citep{ricca10}, therefore, its ratio with the 3.3 $\mu$m PAH could be used as a size indicator. Figure \ref{fig:rat} shows the 17$/$3.3 ratio as a function of 
N$_{\rm C}$ and confirms the ratio as a good tracer of PAH size.
Finally, we examine the effect the
hardness of the underlying radiation field has on the 17$/$3.3 PAH ratio. We
consider two limiting cases with small and large  N$_{\rm c}$  We use the ‘bin-averaged’ values for PAHs at the
low (20$<$N$<$40) and high end (200$<$N$,$400) of the PAH distribution. In Figure \ref{fig:pah_hard} we
show the dependence of the ratio on different values of the hardness of the radiation field and on charge.

\section{The spectral fits}
\label{sec:appB}
\begin{figure*}
	\includegraphics[width=\textwidth]{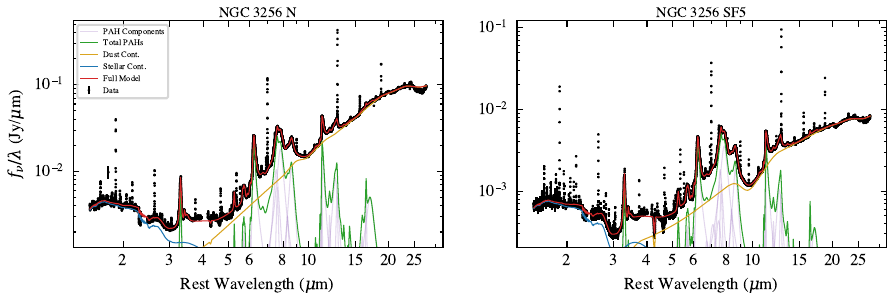}
       \caption{Two examples of fits to NIRSpec and MIRI-MRS spectra
         for the nucleus of NGC 3256 N and the star-forming region NGC
         3256 SF5, using the \citet{fd24} model.}
    \label{fig:example_spectra}
\end{figure*}

The spectra extracted from the apertures shown in Figure \ref{fig:pah-map} were fitted using the model presented in \cite{fd24}. In Figure \ref{fig:example_spectra} we show two examples of such fits for the nucleus of NGC\,3256-N and one of the star-forming regions, NGC\,3256 SF5. The model comprises a dust continuum component (indicated in yellow), a stellar component (blue) and a set of PAH components (purple). The full list of PAH features used in the model can be found in Table B1 of \cite{fd24}. The full model is indicated in red. The spectral fits for the remainder of the apertures shown in Figure \ref{fig:pah-map} can be found in \cite{fd24} and \cite{igb24c}.

\bsp	
\label{lastpage}
\end{document}